\documentclass[journal]{IEEEtran}

\usepackage{amsmath}
\usepackage{amsfonts}

\usepackage[ruled,vlined]{algorithm2e}
\usepackage{subfigure}

\usepackage{array}
\usepackage{bm} 
\usepackage{textcomp}
\usepackage{stfloats}
\usepackage{url}
\usepackage{verbatim}
\usepackage[noadjust]{cite}
\usepackage{xcolor}
\usepackage{amsmath}
\usepackage{amssymb}
\usepackage{mathtools}
\usepackage{graphicx}
\usepackage{epstopdf}
\usepackage{lipsum}

\usepackage{cuted}
\usepackage{amsthm}
\usepackage{soul}

\newtheorem{definition}{Definition}
\newtheorem{assumption}{Assumption}
\newtheorem{lemma}{Lemma}
\newtheorem{theorem}{Theorem}
\newtheorem{corollary}{Corollary}
\newtheorem{proposition}{Proposition}
\newtheorem{observation}{Observation}

\begin{document}
\title{{Diffraction-Aided Wireless Positioning}\\}
\author{
~Gaurav~Duggal,~R.~Michael~Buehrer, ~Harpreet~S.~Dhillon, and Jeffrey~H.~Reed 
\thanks{This article was presented in part at IEEE ICC 2024 ~\cite{duggal20243d} and the ideas in the conference version are subsumed in this work\\
Gaurav Duggal, R. Michael Buehrer, Harpreet S. Dhillon, and Jeffrey H. Reed  are with Wireless@VT,  Bradley Department of Electrical and Computer Engineering, Virginia Tech,  Blacksburg,
VA, 24061, USA. Email: \{gduggal, rbuehrer, hdhillon, reedjh\}@vt.edu.\\ The support of NIST PSCR PSIAP through grant 70NANB22H070, NSF through grant CNS-1923807, CNS-2107276 and NIJ graduate research fellowship through grant 15PNIJ-23-GG-01949-RES is gratefully acknowledged.\\ }
}



\maketitle

\begin{abstract}
Wireless positioning in Non-Line-of-Sight (NLoS) scenarios presents significant challenges due to multipath effects that lead to biased measurements and reduced positioning accuracy. This paper revisits electromagnetic field theory related to diffraction and in the context of wireless positioning and proposes a novel positioning technique that greatly improves accuracy in NLoS environments dominated by diffraction. The method is applied to a critical public safety use case: precisely locating at-risk individuals within buildings, with a particular focus on improving 3D positioning and z-axis accuracy. By leveraging the Geometrical Theory of Diffraction (GTD), the approach introduces an innovative NLoS path length model and a new NLoS positioning technique. Using Fisher information analysis, we establish the conditions required for 3D positioning and derive lower bounds on positioning performance for both 3D and z-axis estimates for the proposed NLOS positioning technique. Additionally, we propose an algorithmic implementation of the proposed NLoS positioning method using non-linear least squares estimation, which we term D-NLS. The positioning performance of our proposed NLoS positioning technique is validated using an extensive ray-tracing simulation. The numerical results highlight the superiority of our approach in outdoor-to-indoor environments, which directly estimates NLoS path lengths and delivers significant performance enhancements over existing methods for both 3D and z-axis positioning scenarios.
\end{abstract}

\begin{IEEEkeywords}
Geometric theory of Diffraction, Time-of-Flight, wireless positioning, z-axis positioning, 3D localization, outdoor-to-indoor propagation, NLoS positioning
\end{IEEEkeywords}

\section{Introduction}
In wireless systems, Time-of-Flight (ToF) measurements are a widely used geometric-based positioning method, commonly employed for position estimation. As a result, this approach is extensively implemented in various positioning technologies, such as GNSS systems \cite{kaplan2017understanding}, WiFi networks, and 4G and 5G networks \cite{dwivedi2021positioning}, among others. In its most general form, we have several anchors with position knowledge and a node at an unknown position, which is to be estimated. ToF measurements are conducted between the anchors and the node to measure the distance between the anchor and the node. We can obtain 2D or 3D position estimates for the node using these measurements. This system works well when Line-of-Sight (LoS) conditions exist when the signal propagation time corresponds to the Euclidean distance between the anchor and the node. However, in Non-Line-Of-Sight (NLoS) scenarios, we face numerous distinct challenges. The wireless signal propagates along paths in 3D space and these are known as Multipath components (MPC) of the signal. The presence of non-resolvable MPCs leads to fluctuations in the received signal strength, and in the case of resolvable MPCs, we can improve the position estimate under certain conditions \cite{shen2010fundamental}. 
In NLoS scenarios where the direct path between the anchor and node does not exist, the NLoS paths are longer than the direct path. ToF-based ranging measurements conducted in this scenario will estimate a path length longer than the direct path, and the excess path
length is termed {\em NLoS bias}. NLoS bias leads to degraded positioning performance and is an important challenge that needs to be resolved, especially for indoor positioning.

\par
Previous work has tackled the problem of NLoS bias using two broad techniques - modeling it either as (a) a stochastic quantity or, (b) as an unknown deterministic quantity that can be estimated under certain conditions. {\em O'Lone et al.} \cite{lonenlosbias} showed mathematically that the NLoS bias due to single-bounce reflections follows the exponential distribution if modeled as a random variable. This approach provides a tractable way to quantify {\em a priori} information associated with the NLoS paths. {\em Shen et al.} \cite{shen2010fundamental} showed that in the absence of {\em a priori} information about the NLoS bias distribution for the NLoS paths, we can discard the NLoS measurements corresponding to these NLoS paths. However, {\em Qi et al.} \cite{qi2006analysis} show that the position estimate improves by incorporating {\em a priori} information about the NLoS bias. This led to the development of algorithms that are robust to NLoS bias \cite{jia2010set,jiacol,venkatesh2007nlos,venkatesh2007non, li2023iterative}.
\par
Moving to the second approach, where the NLoS bias is modeled as a deterministic unknown quantity, {\em Witrisal et al.} \cite{Witrisalet_multipath_indoor_assisted} developed a framework to incorporate information from floor map plans into ToF-based measurements to improve indoor positioning estimates. {\em Leitinger et al.} \cite{leitinger2015evaluation} derived the Cramér-Rao Lower Bound (CRLB) to lower bound the achievable performance for this positioning methodology for the indoor positioning scenario. {\em Mendrzik et al.} \cite{mendrzik_nlos_mpc} assumed that the NLoS paths were generated due to single bounce reflections and proposed to use an antenna array in conjunction with ToF measurements to estimate the point of reflection. This enabled direct estimation of NLoS path lengths, leading to improved position estimates. In the absence of {\em a priori} map information and the ability to directly estimate the incidence points for the NLoS paths, several studies have utilized collaborative techniques \cite{Hassanslam} or proposed SLAM-based approaches \cite{Christianslam, Nazari6DSLAM, leitinger_slam, leitinger2019belief, venus2023graph}. SLAM-based techniques, in particular, leverage the reflection model to simultaneously localize the user and create a map of the unknown environment. This work focuses primarily on the localization aspect, and these SLAM-based approaches could potentially benefit by being extended to include our proposed diffraction-based positioning technique.

A critical use case in modern wireless networks is positioning for public safety scenarios, which poses numerous distinctive challenges. Wireless networks can become highly congested during emergencies such as mass shootings or firefighting incidents, resulting in suboptimal coverage and adversely impacting emergency response efforts. From the industrial perspective, a solution to address this challenge has been devised in the United States through a public-private partnership between the federal government and AT\&T and is called FirstNet \cite{Firstnetroadmap}. To counter network congestion, FirstNet proposed the use of a dedicated band - LTE Band 14 exclusively for public safety usage. In addition, the z-axis position of the UE is measured using a barometer and reported over the dedicated band for positioning purposes. However, this solution relies on fixed cellular base stations, making them susceptible to potential damage, especially during emergencies such as fires, which can result in loss of coverage due to power loss.
\par
Therefore, there is a pressing need for a swiftly deployable private wireless network tailored to meet the exclusive requirements of public safety for both communication and positioning. Such a network would improve indoor coverage and reliability, thus contributing to the development of effective emergency response strategies. Prior art has proposed a system that involves the deployment of mobile UAVs equipped with positional awareness, denoted as `anchors.' These mobile anchors establish wireless connections with User Equipment (UE), called `nodes', located within a building \cite{dwivedi2021positioning, harishetal, duggaletal, duggal20243d, albaneseetal}. Previous analysis of this system demonstrates how this system can improve indoor coverage by leveraging UAV mobility, as discussed in \cite{duggaletal}, while addressing indoor positioning requirements using 5G technology, as elaborated in \cite{harishetal}. However, a key challenge for indoor scenarios lies in further refinement of the accuracy of the z-axis position estimation, especially as quickly navigating between floors poses challenges for emergency responders. The primary source of the position error is due to the presence of NLoS bias in the ToF measurements. 

\par
To summarize, when the nature and number of interactions with objects along the NLOS propagation path of a wireless signal are unknown, we model this path using the Euclidean distance with an additive non-zero stochastic NLOS bias term \cite{lonenlosbias}. Alternatively, we can use Snell's laws to reflect anchors across planar surfaces, leading to {\em virtual} anchors. Now, the NLOS signal propagation path can be modeled using the Euclidean distance from the {\em virtual} anchor, and this is analogous to increasing the number of anchors. However, there are other propagation scenarios in which diffraction from edges might be more common. For example, in Outdoor-to-indoor (O2I) scenarios, diffraction from window edges may actually be more common. 
 Prior works investigating diffraction as a propagation mechanism have been motivated by the need to explain the non-zero electric field at the receiver when the transmitter and receiver have an obstruction in between them. In the absence of a transmitted path and reflections from other obstacles in the environment, a numerical technique to model wireless signal propagation called geometrical optics (GO) incorrectly predicts a zero electric field at the receiver \cite{balanis2012advanced, namara1990introduction}. In this context, diffraction has been studied to model path losses for signal propagation in building scenarios \cite{bostian28, honcharenko1993mechanisms} and for edge imaging applications \cite{anurag_radarconfgtd,pallaprolu2022wiffract}.  In our work, we argue that diffraction from edges plays a dominant role in signal propagation in many scenarios, and thus positioning estimation performance can be improved in NLOS conditions by modeling diffraction in the positioning technique.
 Specifically,  we model the length of the path followed by the diffraction field and incorporate this model in a non-linear least squares positioning technique. We improve on the conference version of this paper \cite{duggal20243d} with a more in-depth inquiry that includes analysis of the diffraction electric field, Fisher information based analysis of the positioning technique and a very comprehensive raytracing simulation to validate our findings. Our main contributions are as follows: 

\begin{itemize}
    \item \textbf{Diffraction-inspired path length model:}
When an obstacle with an edge lies between the transmitter and receiver, it is plausible that the diffraction field propagates from the transmitter to the edge and then to the receiver. However, the precise location of the diffraction point along the edge is not immediately obvious. By applying electromagnetic principles, we derive a closed-form expression to determine the diffraction point's location on the edge. This serves as the fundamental connection between modeling diffraction and wireless positioning and leads to a new diffraction-inspired NLOS path model for positioning.
  
\item \textbf{O2I Signal Propagation Insights}: We argue that diffraction from window edges plays a key role in outdoor-to-indoor (O2I) signal propagation from a transmitter located outside a building to an indoor receiver. To investigate this further we develop a simplified model of a building. Applying the Geometrical Theory of Diffraction (GTD) to this building model, we predict the presence of two diffraction multipath components (MPC's) resulting from diffraction from the upper and lower edges of the building windows. Further, by controlling the transmit side polarization we can control the relative power in the two diffraction MPCs and this leads to a single {\em dominating diffraction-based MPC}. Lastly, we also predict that this {\em dominating diffraction MPC} in the absence of a direct Euclidean path typically is the first arriving path. 
    
 \item \textbf{New NLOS Positioning Technique}: Building on the above results and insights, we propose a new NLOS positioning technique which is applicable to an important public safety scenario where there is a need to improve the indoor positioning performance. This public safety positioning system consists of multiple outdoor UAV anchors and indoor UEs (nodes) that need to be localized. First, we modify the proposed general diffraction path length model to make it more suitable to the O2I signal propagation. This modification improves on the conference version of this paper \cite{duggal20243d} by removing the need to have {\em a priori} knowledge of the exact positions of the diffracting edges. 
Next, we model the range measurements associated with the dominating diffraction MPC using the modified diffraction path length model 
and leverage Fisher information to derive the necessary and sufficient conditions for estimating the 3D position of the indoor node. The Fisher information can also be used to develop the CRLB to lower bound the positioning performance. We also derive a diffraction-based non-linear-least squares positioning algorithm (D-NLS) that relies on the proposed diffraction path length model. Finally, we validate all the developed results using a comprehensive raytracing simulation that models reflections, diffractions and transmissions for this O2I scenario using a popular wireless modeling software \cite{wirelessinsight}. We demonstrate that we can estimate the delay associated with the diffraction paths using the first arriving path principle which leads to greatly improved 3D positioning performance using the D-NLS algorithm. Note that although the diffraction-based path isn't {\it always} the first arriving path, it is very often the first-arriving path.  Also, when it isn't the first-arriving path, the difference in path length between the first-arriving path and the diffraction-based path is small.  This approach is compared both to prior art positioning techniques for NLOS scenarios and also the CRLB. 
\end{itemize}

Our analysis starts by presenting preliminaries from \textbf{electromagnetic} field theory to model diffraction from edges with the goal to obtain the path length and electric field magnitude corresponding to paths resulting from diffraction. We use the results derived in this section to develop a simplified building model with diffraction as the primary propagation mechanism, based on which we develop a ToF based positioning technique. Consequently, we develop a Fisher information framework to analyze the proposed positioning technique and conclude the paper by presenting a public safety scenario where we show improvement in both the 3D and z-axis positioning performance.

\section{Asymptotic Techniques To Model Diffraction}
In this section, we present some preliminaries from electromagnetic field theory. Using the Geometrical Theory of Diffraction (GTD)  \cite{keller1962geometrical,namara1990introduction,balanis2012advanced}, we first present a canonical example known as {\em 3D edge diffraction for a half plane} to model diffraction from edges.
Rather than numerically evaluating Maxwell's equations, classical Geometrical Optics (GO) models refraction and reflection phenomena using ray propagation governed by simple geometrical rules. GTD extends GO, enabling the modeling of diffraction phenomena through ray propagation with similar straightforward geometrical principles.  Another asymptotic technique to model diffraction is the Knife Edge Diffraction (KED) which in contrast to GTD does not include the electric field polarization aspects. Previously, GTD has been extensively adopted to model diffraction in different scenarios concerning edges \cite{anurag_radarconfgtd, pallaprolu2022wiffract, honcharenko1993mechanisms, bostian28,pathak_gtd}. 
\subsection{3D Edge Diffraction for a Half Plane}\label{section_3D_edge_diffraction_half_plane}
\begin{figure*}[hpt]
    \centering
    \subfigure[3D Edge diffraction for a half plane.]{ 
        \begin{minipage}[t]{0.49\textwidth}
        \centering
        \includegraphics[ width=\linewidth]{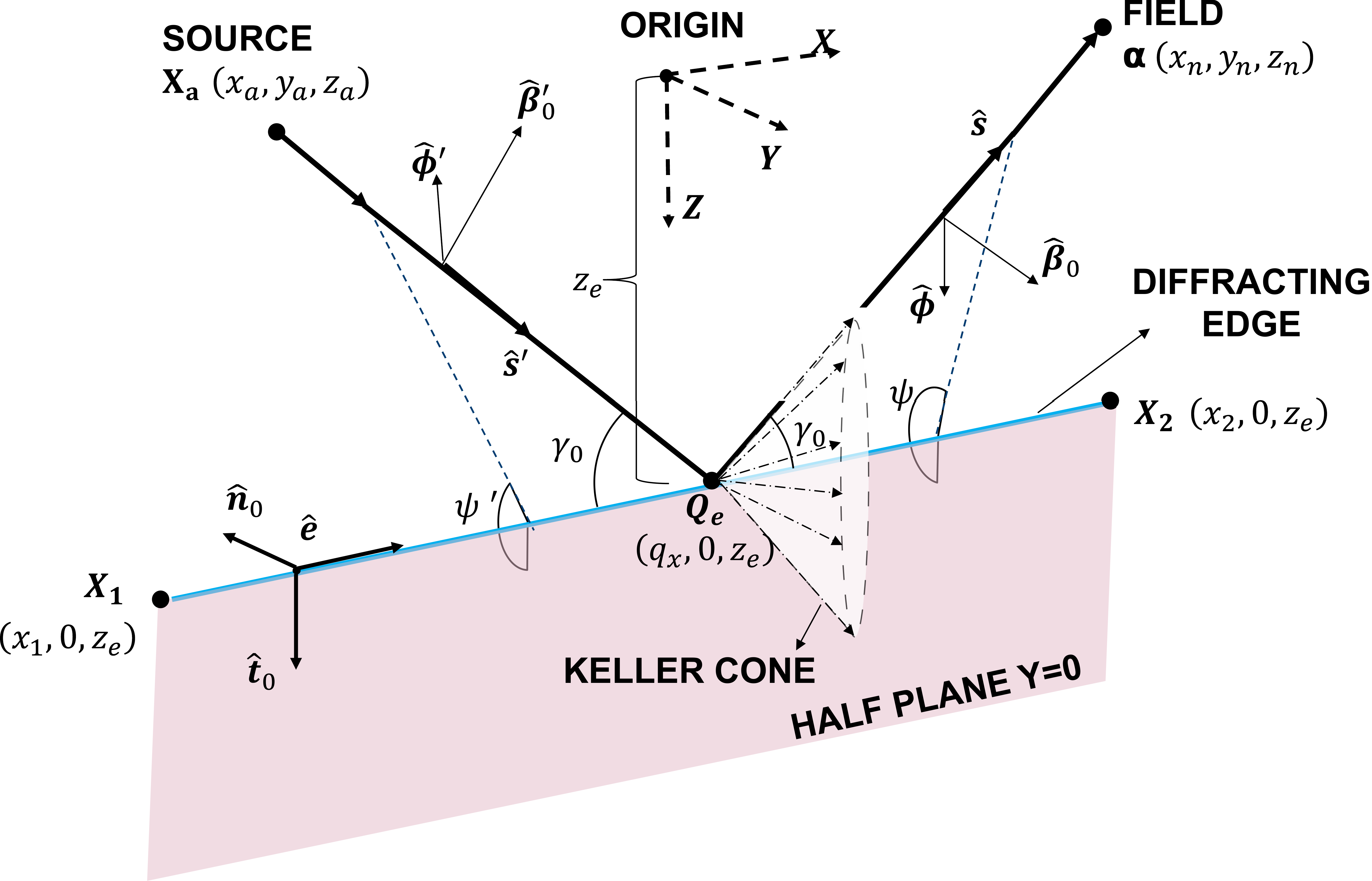}
        \label{fig_GTD_half_plane}    
        \end{minipage}}
        \subfigure[Simplified model of the building.]{
        \begin{minipage}[t]{0.49\textwidth}
        \centering
        \includegraphics[ width=\linewidth]{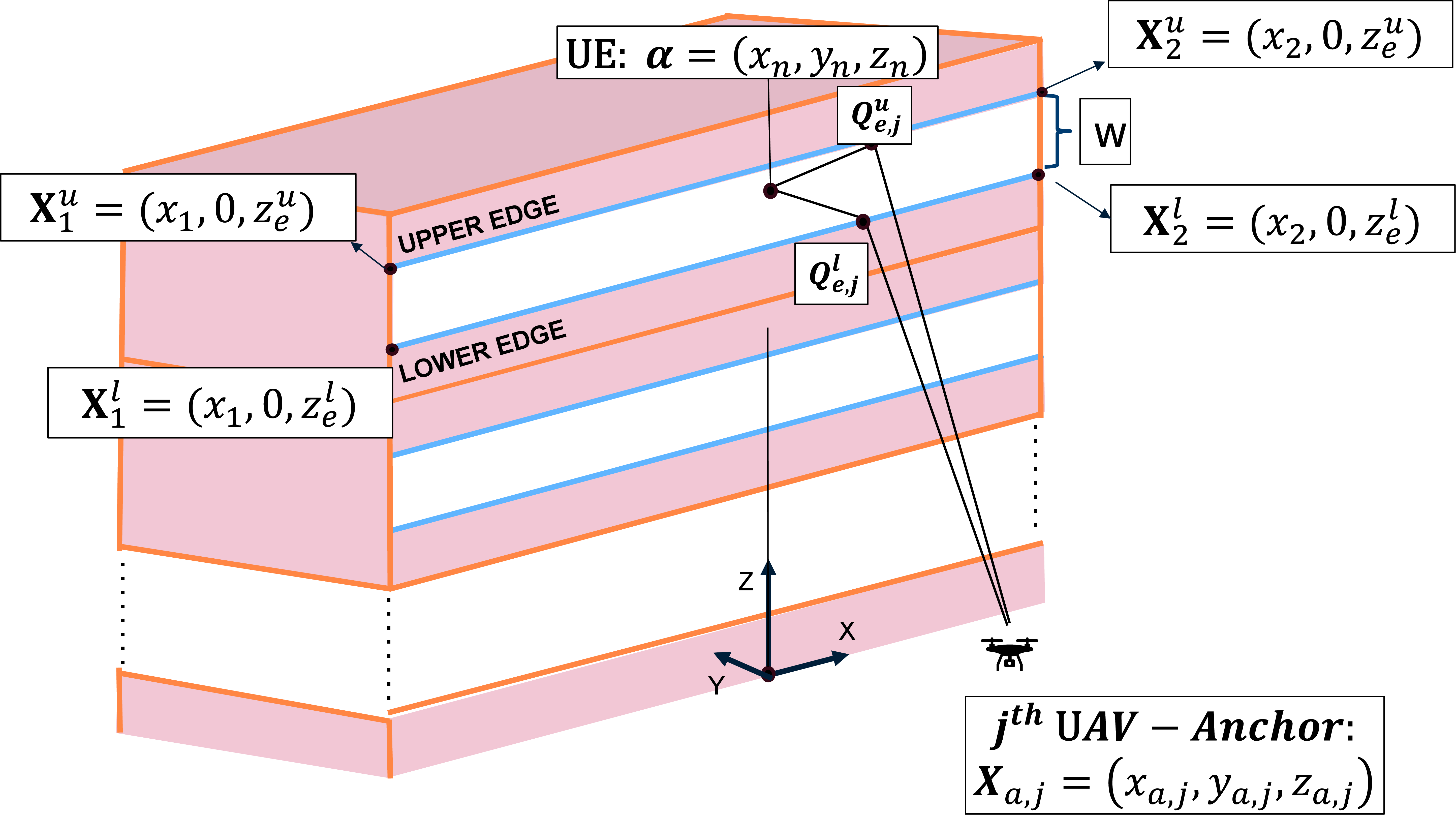}
        \label{fig_simplified_building_model}  
        \end{minipage}}

        \caption{Application of the Geometrical Theory of Diffraction (GTD) (a) Canonical example to model edge diffraction in 3D for a half plane (b) Application of the half plane model to an O2I signal propagation with the transmitter on the UAV-anchor and receiver inside the building.} 
        \label{fig:GTD_geometry}
 \end{figure*}

In this section, we aim to derive closed form expressions for both the path length and the electric field strength of the signal propagation due to diffraction from an edge. 
Consider Fig.\,\ref{fig_GTD_half_plane}; we have a perfectly conducting half-plane at $Y=0$ whose edge is a distance $z_e$ from the origin and parallel to the X-axis. Note that the edge itself is considered to be infinite in length; however, in practice, a finite length corresponding to several wavelengths is sufficient. The source is at point $\bm{X}_a$, the observation is at point $\bm{\alpha}$, and both these points are located on different sides of the half-plane. The path the diffraction electric field follows is along ray $\bm{X_a}\bm{Q_e}$ and then $\bm{Q_e}\bm{\alpha}$ such that the point $\bm{Q_e}$ lies on the diffracting edge. The location of the point $\bm{Q_e}$ on the edge is such that the geometrical distance $\bm{X_aQ_e\alpha}$ is minimized. This is known as Fermat's principle of least time \cite{born2013principles}. Therefore, the location of the diffraction point $\bm{Q_e}$ is ultimately dependent on both the relative position of the source and observation point with respect to the diffracting edge. Fermat's principle can also be shown equivalent to the diffraction law \cite{namara1990introduction, balanis2012advanced}, and we use this to derive a closed-form expression of the path length specified by the geometrical distance $\bm{X_aQ_e\alpha}$. The diffraction law predicts that the incident ray on an edge leads to a cone of diffracted rays. This cone has been observed under certain conditions \cite{rahmat2007keller} and is referred to as the Keller cone. In terms of the diffraction law, the location of the tip of the Keller cone $\bm{Q_e}$ is dependent on the relative location of the source point $\bm{X_a}$ and observation point $\bm{\alpha}$ with respect to the edge. Next, we choose the diffracted ray on this Keller cone that passes through the observation point $\bm{\alpha}$. Hence, $\bm{X_aQ_e\alpha}$ represents the path followed by the diffracted field to reach point $\bm{\alpha}$. 
\par
Now, we express the diffraction law in more mathematical terms by defining the edge vector $\bm{\hat{e}}$ to point in the tangential direction to the edge. To resolve the ambiguity of the two possible tangential directions, we define two other vectors -- $\bm{\hat{n}_0}$ as normal to the half plane in the direction of the source point $\bm{X_a}$ and $\bm{\hat{t}_0}$ as the normal to the edge vector and pointing away from the edge along the surface of the half-plane. These three orthonormal vectors are shown in Fig.\,\ref{fig_GTD_half_plane} and satisfy
 \begin{equation}
     \bm{\hat{t}}_0 = \bm{\hat{n}}_0 \times \bm{\hat{e}}.
 \end{equation}

\begin{definition}
\label{definition_law_of_diffraction}
The law of diffraction states that the angle $\gamma_0$ between the incident ray $\bm{\hat{s}^{'}}$ and the edge $\bm{\hat{e}}$ is the same angle between the diffracted ray $\bm{\hat{s}}$ and the edge $\bm{\hat{e}}$. This is mathematically expressed as
\begin{equation}
\cos \gamma_0= \bm{\hat{s}^{\prime}} \cdot \bm{\hat{e}} = \bm{\hat{s}} \cdot \bm{\hat{e}}, \\ \;\;\;\;\;\;  0 \le \gamma_0 \le \frac{\pi}{2}.
\label{eq_law_of_diffraction}
\end{equation}
Note, here $\gamma_0$ is the half angle of the Keller cone as shown in Fig.\,\ref{fig_GTD_half_plane}. 
\end{definition}
Here $\bm{\hat{s}^{\prime}}$, $\bm{\hat{s}}$ and $\bm{\hat{e}}$ are the unit vectors along the incident ray, the diffracted ray and the edge, respectively, and $\overrightarrow{\bm{A}} \cdot \overrightarrow{\bm{B}}$ denotes the dot product between the vectors $\overrightarrow{\bm{A}}$ and $\overrightarrow{\bm{B}}$. 
\subsection{Path Length Model}
In this section, we apply the law of diffraction to derive a closed-form expression for the path length $\bm{X_aQ_e\alpha}$, which is presented in the next lemma.
\begin{lemma}
\label{lemma_half_plane_path_length}
The path length $p$ traced out by the geometrical path $\bm{X_aQ_e\alpha}$ in Fig.\,\ref{fig_GTD_half_plane} is expressed as 
\begin{equation}
\label{eq_path_length_half_plane}
\begin{split}
p = & \sqrt{(x_{a}-q_x)^2+(y_{a})^2+(z_{a}-z_{e})^2} \\ 
&+  \sqrt{(x_n-q_x)^2+(y_n)^2+(z_e-z_n)^2}.\\
\end{split}
\end{equation}
Here, the coordinates of the diffraction point $\bm{Q}_e=[q_x,0,z_e]^T$, can be expressed as a convex combination of the coordinates of the endpoints of the edge $\bm{X}_1=[x_1,0,z_e]^T$ and $\bm{X}_2=[x_2,0,z_e]^T$ as

\begin{equation}
\begin{split}
\bm{Q}_e & = \lambda \bm{X}_1 + (1-\lambda) \bm{X}_2,\;
\lambda  = \frac{-b \pm \sqrt{b^2-4ac}}{2a},  \\
a &  = (x_1-x_2)^2 \left[(y_n^2-y_{a}^2)+ (z_{n}^2-z_a^2)  +2z_e(z_a-z_{n})\right], \\
b &  = 2(x_1-x_2) \left[(x_2-x_a)\left((z_e-z_{n})^2+y_n^2\right) \right. \\& \left. \;\;\;\;\;\;\;\;\;\;\;\;\;\;\;\;\;\;\;\;\;\;\;\;\;\;\;\;\; -(x_2-x_n)  \left((z_e-z_a)^2+y_a^2\right)\right], \\
c & = (x_2-x_a)^2 \left[(z_e-z_{n})^2+y_n^2\right] \\&\;\;\;\;\;\;\;\;\;\;\;\;\;\;\;\;\;\;\;\;\;\;\;\;\;\;\;\;\;   - (x_2-x_n)^2  \left[(z_e-z_a)^2+y_a^2\right].
\label{eq_quadratic_qe}
\end{split}
\end{equation}
\end{lemma}
\begin{IEEEproof}
    The proof uses Definition \ref{definition_law_of_diffraction} and is presented in Appendix \ref{derivation_path_length}.
\end{IEEEproof}
This result can also be derived from Fermat's principle of least time. By applying this principle, we calculate the derivative of the path length expression in eq. \eqref{eq_path_length_half_plane} with respect to the x-coordinate $q_x$ of the diffraction point $\bm{Q_e}=[q_x,0,z_e^u]^T $ and set the expression obtained to zero, that is, $\frac{\delta p}{\delta q_x}=0$. This gives the same expression as is obtained from the diffraction law. 
\par
This is a general path length expression applicable for modeling diffraction from any edge. In the conference version of this article \cite{duggal20243d}, we used this expression for indoor positioning; however, it required detailed positional knowledge of the diffracting edges. Corollary \ref{corollary_diffraction_path_length_building} in section \ref{sec_simplified_model_of_building} presents an improved diffraction path length expression tailored to the modeled scenario, and in Corollary \ref{corollary_conditions_for_positioning} in section \ref{section_FIM_derivation} we derive the necessary and sufficient conditions for positioning using this model. Thus, we are able to eliminate the need for detailed positioning knowledge about the diffraction edges.

\subsection{Diffraction Electric Field}
We derive an expression for the received diffraction electric field at the observation point $\bm{\alpha}$ in Fig.\,\ref{fig_GTD_half_plane}. The electric field polarization along the incident ray is resolved into two perpendicular components $\bm{\hat{\beta}_0^{\prime}}$, $\bm{\hat{\phi}^{\prime}}$ both of which are also perpendicular to the propagation direction of the incident ray $\bm{\hat{s^\prime}}$. Similarly, the electric field of the diffracted ray is resolved into $\bm{\hat{\bm{\beta}}_0}$, $\bm{\hat{\phi}}$ components, both perpendicular to the propagation direction of the diffracted ray $\bm{\hat{s}}$. We now define additional vectors to define two new coordinate systems, called the ray fixed coordinate system and the edge fixed coordinate system.


\begin{equation}
\begin{split}
\bm{\hat{\phi}^{\prime}}  =\frac{-\bm{\hat{e}} \times \bm{\hat{s}^{\prime}}}{\left|\bm{\hat{e}} \times \bm{\hat{s}^{\prime}}\right|}, \; 
\bm{\hat{\beta}_0^{\prime}}  = \bm{\hat{\phi}^{\prime}} \times \bm{\hat{s}^{\prime}},\; 
\bm{\hat{\phi}}  =\frac{\bm{\hat{e}} \times \bm{\hat{s}}}{|\bm{\hat{e}} \times \bm{\hat{s}}|}, \;
\bm{\hat{\beta}_0}  = \bm{\hat{\phi}} \times \bm{\hat{s}}.
\end{split}
\end{equation}

\begin{definition}
\label{definition_electric_field_half_plane}
The diffracted field $\bm{\overrightarrow{E}^d}$ due to the edge $\bm{\hat{e}}$ of the half-plane can be expressed along two orthogonal vectors $\bm{\hat{\beta}_{0}}$ and $\bm{\hat{\phi}}$ as
\begin{align}
& \bm{\overrightarrow{E}^d}=E_{\beta_0}^d \bm{\hat{\beta}_0}+E_{\hat{\phi}}^d \bm{\hat{\phi}}.
\label{eq_total_ef}
\end{align}
\end{definition}

Here, $E_{\beta_{0}}^{d}$ and $E_{\phi}^{d}$ are the diffracted electric field components along the $\bm{\hat{\beta}_{0}}$ and $\bm{\hat{\phi}}$ directions and are obtained using 

\begin{equation}
\left[\begin{array}{c}{E}_{\beta_0}^d \\ {E}_{\phi}^d\end{array}\right]=\left[\begin{array}{cc}-D_s & 0 \\ 0 & -D_h\end{array}\right]\left[\begin{array}{c}{E}_{{\beta _0}^{\prime}}^i\left(\bm{Q_e}\right) \\ {E}_{\phi^{\prime}}^i\left(\bm{Q_e}\right)\end{array}\right] \frac{\mathrm{e}^{-j k |\bm{s}|} }{\sqrt{|\bm{s}|}}.
\label{eq_GTD_ef_edge}
\end{equation}
Here, $D_{s}$ and $D_{h}$ are the soft and hard diffraction coefficients, respectively, $|\bm{s}|$ is the distance between the field point $\bm{\alpha}$ and the diffraction point $\bm{Q_e}$ and $k$ is the wavenumber. ${E}_{{\beta _0}^{\prime}}^i\left(\bm{Q_e}\right)$ and ${E}_{\phi^{\prime}}^i\left(\bm{Q_e}\right)$ are the components of the incident electric field at the point $\bm{Q_e}$ along $\bm{\hat{\beta}_0^{\prime}}$ and $\bm{\hat{\phi}^{\prime}}$ directions respectively. The incident field components at point $\bm{Q_e}$ are calculated using


\begin{equation}
\begin{split}
  {E}_{{\beta _0}^{\prime}}^i\left(\bm{Q_e}\right) = \left( \bm{\overrightarrow{E}_0} \cdot \bm{\hat{\beta}_0^{\prime}} \right)  \frac{\mathrm{e}^{-j k |\bm{s^{\prime}}|} }{|\bm{s^{\prime}}|} ,\\
  {E}_{\phi^{\prime}}^i\left(\bm{Q_e}\right) = \left( \bm{\overrightarrow{E}_0} \cdot \bm{\hat{\phi}^{\prime}} \right)  \frac{\mathrm{e}^{-j k |\bm{s^{\prime}}|} }{|\bm{s^{\prime}}|}. 
  \end{split}
\label{eq_incident_ef_qe_perpendicular_beta_0_phi}
\end{equation}

Here, $\bm{\overrightarrow{E}_0}$ is the linearly polarized transmitted electric field polarization vector at the source point $\bm{X_a}$, $\bm{|s^{\prime}}|$ is the distance from the source point $\bm{X_a}$ to the diffraction point $\bm{Q_e}$.
Finally, the soft ($D_s$) and hard ($D_h$) Keller Diffraction coefficients for a thin sheet are given by \cite{namara1990introduction,balanis2012advanced}
\begin{equation}
\begin{split}
    D_{s}=\frac{-e^{-j \pi / 4}}{2 \sqrt{2 \pi k} \sin \gamma_0}\left[\frac{1}{\cos \left(\frac{\psi-\psi^{\prime}}{2}\right)} - \frac{1}{\cos \left(\frac{\psi+\psi^{\prime}}{2}\right)}\right], \\
    D_{h}=\frac{-e^{-j \pi / 4}}{2 \sqrt{2 \pi k} \sin \gamma_0}\left[\frac{1}{\cos \left(\frac{\psi-\psi^{\prime}}{2}\right)} + \frac{1}{\cos \left(\frac{\psi+\psi^{\prime}}{2}\right)}\right].
\end{split}
\label{eq_diffraction_coeff}
\end{equation}
The angles $\psi^{\prime}$ and $\psi$ are obtained using the procedure below. We first define unit vectors $\bm{\hat{s}_t^{\prime}}$ and $\bm{\hat{s}_t}$ as


\begin{equation}
\begin{split}
\bm{\hat{s}_t^{\prime}}=\frac{\bm{\hat{s}^{\prime}}-\left(\bm{\hat{s}^{\prime}} \cdot \bm{\hat{e}}\right) \bm{\hat{e}}}{\left|\bm{\hat{s}^{\prime}}-\left(\bm{\hat{s}^{\prime}} \cdot \bm{\hat{e}}\right) \bm{\hat{e}}\right|},\;\; 
\bm{\hat{s}_t}=\frac{\bm{\hat{s}}-(\bm{\hat{s}} \cdot \bm{\hat{e}}) \bm{\hat{e}}}{\left|\bm{\hat{s}}-(\bm{\hat{s}} \cdot \bm{\hat{e}}) \bm{\hat{e}}\right|},
\end{split}
\label{eq_s_t_prime_and_s_t}
\end{equation}

to lie in a plane perpendicular to the edge. Then the angles ${\psi}^{\prime}$ and ${\psi}$ are given by
\begin{equation}
\begin{split}
\psi^{\prime}&=\pi-\left[\pi-\arccos \left(-\bm{\hat{s}_t^{\prime}} \cdot \bm{\hat{t}_o}\right)\right] \operatorname{sgn}\left(-\bm{\hat{s}_t^{\prime}} \cdot \bm{\hat{n}_o}\right), \\
\psi&=\pi-\left[\pi-\arccos \left(\bm{\hat{s}_t} \cdot \bm{\hat{t}_o}\right)\right] \operatorname{sgn}\left(\bm{\hat{s}_t} \cdot \bm{\hat{n}_o}\right).
\end{split}
\label{eq_psi_prime_psi}
\end{equation}
Note, $ 0 \le \arccos{x} \le \pi$ and the function $\operatorname{sgn}\left(x\right)$ refers to the standard signum function definition. These diffraction coefficients were initially proposed by Keller \cite{keller1962geometrical} and are valid for regions away from the shadow boundaries. However, they are discontinuous over the shadow boundaries and, to resolve the discontinuity, they can be replaced with the UTD diffraction coefficients \cite{kouyoumjian1974uniform}. Since the UTD and GTD coefficients agree for regions away from the shadow boundaries \cite{balanis2012advanced}, this assumption does not affect our results.

\section{Application to an O2I Scenario}
\label{section_application_to_O2I}

\begin{figure}[t!]
\centering
     \includegraphics[width=0.6\linewidth]{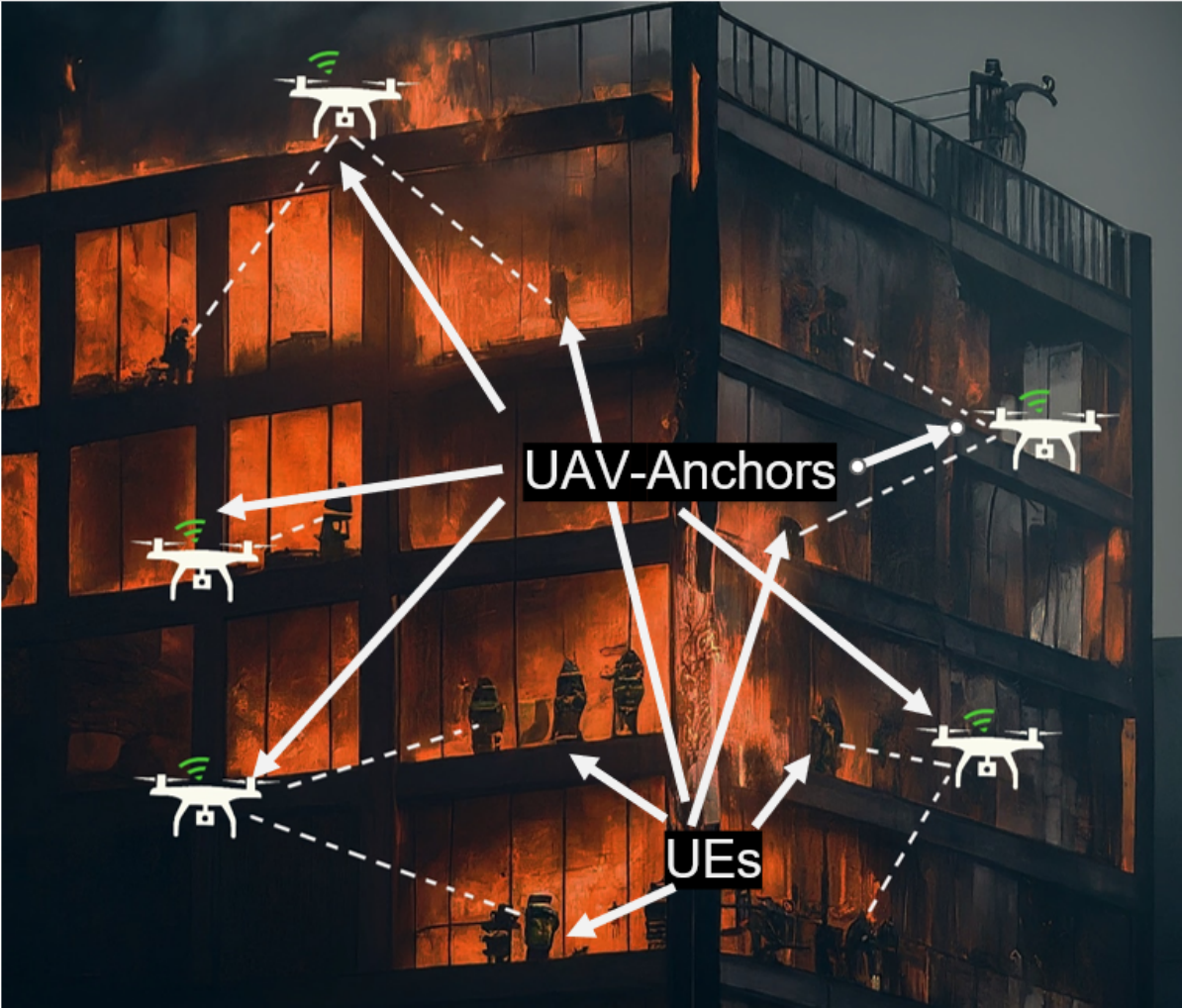}
         \caption{Positioning System consisting of UAV anchors with position knowledge connected to indoor UEs carried by at-risk individuals within the building}
         \label{fig_positioning_system}
\end{figure}

Consider Fig.\,\ref{fig_positioning_system} for an emergency scenario with emergency responders at unknown locations inside a multi-floored building. We have a positioning system consisting of UAVs that are quickly deployed around the affected building. These UAVs form a wireless network, have position knowledge, and behave as anchors at known locations. This represents an O2I signal propagation scenario. {\em Kohli et al.} \cite{kohlio2i} conducted a path loss-based analysis at 28 GHz for the O2I scenario and concluded that the O2I path loss depended on the type of glass on the exterior of the building.
{\em Bas et al.} \cite{bas2019outdoor} conducted a measurement campaign for a similar O2I scenario in which they investigated the propagation of the signal from an outdoor transmitter to an indoor receiver. They concluded that at 28 GHz for brick buildings, the direct path between the transmitter and receiver if blocked by the exterior of the building is severely attenuated. In addition, they concluded that the MPCs observed in indoor locations were the result of interactions with windows. Our analysis of the diffraction paths in Section \ref{section_3D_edge_diffraction_half_plane} based on Fermat's principle of least time suggests that the first arriving paths at indoor locations are the result of diffraction from the window edges. Furthermore, if we can estimate the path lengths of these diffraction paths, we could circumvent the problem of NLoS bias, thereby improving both the 3D and z-axis position estimates.   

Our analysis starts by developing a simplified model of the building to model the diffraction MPCs. Based on this model, we develop insight into the propagation of the signal from the outdoor anchor to the indoor node. We also present results from realistic electromagnetic modeling software that uses raytracing to investigate different signal propagation mechanisms and validate some of our insights.   

\subsection{Simplified Model of a Building}\label{sec_simplified_model_of_building}
 Our simplified model of the building begins by assuming that every floor of the building can be modeled as two half-planes separated by a vertical distance $w$ corresponding to the vertical dimension of the window, as shown in Fig.\ref{fig_simplified_building_model}. Consider a node that is located on an arbitrary floor of the building at $\bm{\alpha}=[x_n,y_n,z_{n}]^T$. The transmit signal originates from the $j^{\text{th}}$ anchor at a known location $\bm{X}_{a,j}=[x_{a,j},y_{a,j},z_{a,j}]^T$ below the building floor where the node is located, such that there is NLoS signal propagation. At the node location, we would obtain two diffraction MPCs resulting from the diffraction from the upper and lower edges of the window located on the same floor of the building. Note that our model does not contain vertical pillars that separate windows on the same floor. This is because, applying the lemma \ref{lemma_half_plane_path_length} for the vertical edges located on the vertical pillars, we would obtain $\lambda>1$ or $\lambda<0$ when solving the corresponding quadratic equation. This implies that the diffraction point $\bm{Q_e}$ lies beyond the endpoints of the vertical edges. In this case, the diffraction would be from the corner where the window's horizontal and vertical edges meet. This phenomenon is called {\em corner diffraction}. In our work, we ignore it because it is known to be significantly weaker compared to edge diffraction \cite{siktaetal}.  
\par
An important observation is that the diffraction MPCs are generated from the edges of the window located on the same floor as the node. This suggests that z-axis position information is inherently encoded in the diffraction MPCs. Our goal is to obtain the path length and the electric field for these two diffraction MPCs from the upper and lower window edges. We use superscripts `$u$' and `$l$' to refer to the upper and lower diffraction MPC parameters, respectively. 
\begin{assumption}
\label{assumption1_diffraction_same_floor}
In our simplified model of a building shown in Fig.\,\ref{fig_simplified_building_model}, the active diffraction edges are located on the same building floor as the node. We assume that the z-coordinate of the node and the diffracting edges are always separated by a constant offset. Without loss of generality, we assume that the vertical coordinate of the node $z_{n}$ is located at the midpoint of the node building floor and that the window center coincides with the midpoint of the node building floor. Therefore, $z_{n}=z_e^u-0.5w$ for the upper edge diffraction MPC and $z_{n}=z_e^l+0.5w$ for the lower edge diffraction MPC. Here, $z_e^u$ and $z_e^l$ are the vertical coordinates of the upper and lower diffracting edges, respectively and $w$ is the vertical dimension of the window.
\end{assumption}
The distance between the node's z-coordinate and the diffracting edges generally depends on the receiver's mounting height above the building floor, the vertical separation of the diffracting edges, and their relative positions with respect to the node. This assumption is used only to arrive at the analytical results. We discuss the impact of relaxing this assumption (D-NLS(Case-1) vs D-NLS (Case-2)) for our proposed positioning technique in the numerical results in Section \ref{section_non_linear_least_squares}. This is done by introducing a model mismatch using a vertical offset $\Delta$ in eq. \eqref{eq_assumption1_vertical_offset}. 
\begin{corollary}
\label{corollary_diffraction_path_length_building}
Two diffraction MPCs originate from an anchor location $\bm{X_{a,j}}$ outside the building to a node location $\bm{\alpha}$ on an arbitrary floor inside the building. These two MPCs are due to diffraction from the lower and upper edges of the window with vertical dimensions $w$. This window is situated on the same building floor as the node. For each MPC, assuming that the diffraction point $\bm{Q_e}$ lies on the respective diffracting edge, the path length can be expressed as the sum of two Euclidean distances $|\bm{X_{a,j}Q_e}|$ as the Outside Path Length (OPL) and $|\bm{Q_e\alpha}|$ as the Inside Path Length (IPL). The path length expression for the upper edge MPC is
\begin{equation}
\label{eq_path_length_building}
\begin{split}
p_{j}(\bm{\alpha}) = & \underbrace{\sqrt{(x_{a,j}-q_{x,j}^u)^2+(y_{a,j})^2+(z_{a,j}-z_{n}-0.5w)^2}}_{OPL} \\ 
&+  \underbrace{\sqrt{(x_n-q_{x,j}^u)^2+(y_n)^2+(0.5w)^2}}_{IPL}.\\
\end{split}
\end{equation}
Here, the diffraction point coordinates for the upper edge diffraction path are $\bm{Q_e^u}=[q_x^u,0,z_{n}+0.5w]^T$. Note that $q_x^u$ can be obtained using eq. \eqref{eq_quadratic_qe}.
\end{corollary}
\begin{IEEEproof}
 We substitute the value of the z-coordinate of the upper edge in the Assumption \ref{assumption1_diffraction_same_floor} into the expression of the path length in Lemma \ref{lemma_half_plane_path_length} to obtain a closed-form expression for the path length of the upper edge diffraction MPC. Note that we need to substitute the appropriate coordinates of the end points $\bm{X}_1^u$ and $\bm{X}_2^u$ of the upper diffracting edge to calculate the corresponding parameter $\lambda$ to obtain the path length. With similar arguments, we can also get the expression for the path length of the diffraction MPC due to the lower diffraction edge.
\end{IEEEproof}

We now derive the electric field expressions for the two diffraction MPCs. The electric field for each edge can be calculated using the electric field expression for a half-plane presented in Definition \ref{definition_electric_field_half_plane}. For each edge, we obtain expressions for the incident ray vector $\bm{\hat{s}^{'}}$, diffracted ray vector $\bm{\hat{s}}$, and the edge vector $\bm{\hat{e}}$ and the final expressions are presented in Table \ref{Table_edge_incident_diffracted_ray}. 
\begin{table*}[!t]
\caption{Parameters for upper and lower edges for $\bm{\hat{e}}$, $\bm{\hat{s}^{'}}$ and $\bm{\hat{s}}$.}
\label{Table_edge_incident_diffracted_ray}
\centering
\begin{tabular}{|c|c|c|c|c|c|c|}
\hline
 & $\bm{Q_e}$ & $\bm{\hat{n}_0}$ & $\bm{\hat{t}_0}$ & $\bm{\hat{e}}$ & $ \bm{\hat{s}^\prime}$ & $\bm{\hat{s}}$\\

\hline
Lower edge & $q_x^l\bm{\hat{x}}+0\bm{\hat{y}}+(z_e^l)\bm{\hat{z}}$ & $-\bm{\hat{y}}$ & $-\bm{\hat{z}}$ & 
$-\bm{\hat{x}}$ & $\frac{(q_x^l-x_a)\bm{\hat{x}}-(y_a)\bm{\hat{y}}+(z_e^l-z_a)\bm{\hat{z}}}{\sqrt{(q_x^l-x_a)^2+(y_a)^2+(z_e^l-z_a)^2}}$ & $\frac{(x_n-q_x^l)\bm{\hat{x}}+(y_n)\bm{\hat{y}}+(0.5w)\bm{\hat{z}}}{\sqrt{(x_n-q_x^l)^2+(y_n)^2+(0.5w)^2}}$ \\

\hline
Upper edge & $q_x^u\bm{\hat{x}}+0\bm{\hat{y}}+(z_e^u)\bm{\hat{z}}$& $-\bm{\hat{y}}$ & $\bm{\hat{z}}$& $\bm{\hat{x}}$ & $\frac{(q_x^u-x_a)\bm{\hat{x}}-(y_a)\bm{\hat{y}}+(z_e^u-z_a)\bm{\hat{z}}}{\sqrt{(q_x^u-x_a)^2+(y_a)^2+(z_e^u-z_a)^2}}$ & $\frac{(x_n-q_x^u)\bm{\hat{x}}+(y_n)\bm{\hat{y}}-(0.5w)\bm{\hat{z}}}{\sqrt{(x_n-q_x^u)^2+(y_n)^2+(0.5w)^2}}$\\

\hline
\end{tabular}
\end{table*}

\begin{table*}[!t]
\caption{Edge coordinates for upper and lower edges.}
\label{Table_edge_coord_vectors_edges}
\centering
\begin{tabular}{|c|c|c|}
\hline
Params & Lower Edge &  Upper Edge  \\
\hline
$\bm{\hat{\phi}^\prime}$ & $\frac{(z_e^l-z_a)\bm{\hat{y}}-(y_a)\bm{\hat{z}}}{\sqrt{(y_a)^2+(z_e^l-z_a)^2}}$  & $\frac{(z_e^u-z_a)\bm{\hat{y}}+(y_a)\bm{\hat{z}}}{\sqrt{(y_a)^2+(z_e^u-z_a)^2}}$    \\
\hline
$\bm{\hat{\bm{\beta}}^{\prime}_0}$ & $\frac{-(y_a^2+(z_e^l-z_a)^2)\bm{\hat{x}}-(y_a)(q_x^l-x_a)\bm{\hat{y}}+(z_e^l-z_a)(q_x^l-x_a)\bm{\hat{z}}}{\sqrt{(y_a)^2+(z_e^l-z_a)^2}\sqrt{(q_x^l-x_a)^2+(y_a)^2+(z_e^l-z_a)^2}}$ & $\frac{(y_a^2+(z_e^u-z_a)^2)\bm{\hat{x}}+(y_a)(q_x^u-x_a)\bm{\hat{y}}-(q_x^u-x_a)(z_e^u-z_a)\bm{\hat{z}}}{\sqrt{(y_a)^2+(z_e^u-z_a)^2}\sqrt{(q_x^u-x_a)^2+(y_a)^2+(z_e^u-z_a)^2}}$ \\
\hline
$\bm{\hat{\bm{\phi}}}$ & $\frac{(0.5w)\bm{\hat{y}}-(y_n)\bm{\hat{z}}}{\sqrt{(y_n)^2+(0.5w)^2}}$ & $\frac{(0.5w)\bm{\hat{y}}+(y_n)\bm{\hat{z}}}{\sqrt{(y_n)^2+(0.5w)^2}}$\\
\hline
 $\bm{\hat{\bm{\beta}_{0}}}$ & $\frac{(y_n^2+(0.5w)^2)\bm{\hat{x}}-(y_n)(x_n-q_x^l)\bm{\hat{y}}+(0.5w)(x_n-q_x^l)\bm{\hat{z}}}{\sqrt{(y_n)^2+(0.5w)^2}\sqrt{(x_n-q_x^l)^2+(y_n)^2+(0.5w)^2}} $ & $\frac{-(y_n^2+(0.5w)^2)\bm{\hat{x}}+(y_n)(x_n-q_x^u)\bm{\hat{y}}-(0.5w)(x_n-q_x^u)\bm{\hat{z}}}{\sqrt{(y_n)^2+(0.5w)^2}\sqrt{(x_n-q_x^u)^2+(y_n)^2+(0.5w)^2}}$ \\
\hline
\end{tabular}
\end{table*}

Assume that the anchor transmits a linearly polarized electric field $\bm{E_0} = [E_{0x},E_{0y},E_{0z}]^T$, for each edge, the incident field at the diffraction point $\bm{Q_e}$ lying on the respective edge is expressed as a vector with components along orthogonal vectors $\bm{\hat{\bm{\beta}}_0^\prime}$ and $\bm{\hat{\bm{\phi}}^\prime}$. These orthogonal vectors are calculated using eq. \eqref{eq_incident_ef_qe_perpendicular_beta_0_phi} for both edges.
Similarly, the diffracted electric field for each edge is obtained along two orthogonal vectors $\bm{\hat{\bm{\beta}}_0}$ and $\bm{\hat{\bm{\phi}}}$ using eq. \eqref{eq_GTD_ef_edge}.
Let the incident field orthogonal vectors be $\bm{\hat{\bm{\beta}}_0^\prime} = [\hat{\bm{\beta}}_{0x}^\prime, \hat{\bm{\beta}}_{0y}^\prime, \hat{\bm{\beta}}_{0z}^\prime]^T$, $\bm{\hat{\bm{\phi}}^\prime} = [\hat{\bm{\phi}_x}^\prime,\hat{\bm{\phi}_y}^\prime,\hat{\bm{\phi}_z}^\prime]^T$ and the diffracted field orthogonal vectors be $\bm{\hat{\bm{\beta}}_0} = [\hat{\bm{\beta}}_{0x},\hat{\bm{\beta}}_{0y},\hat{\bm{\beta}}_{0z}]^T$, $\bm{\hat{\bm{\phi}}} = [\hat{\bm{\phi}_x},\hat{\bm{\phi}_y},\hat{\bm{\phi}_z}]^T$ where the subscript `$x$',`$y$' and `$z$' represents the component along the X,Y and Z axes, respectively. 

\begin{corollary}
\label{corollary_diffraction_ef_building}
The electric field corresponding to each of the diffraction MPCs in our simplified building model is given by
\begin{equation}
\begin{split}
\bm{E^d} & = \left[\left(E_{0x}\hat{\beta}_{0x}^\prime +  E_{0y}\hat{\beta}_{0y}^\prime + E_{0z}\hat{\beta}_{0z}^\prime\right) \right. \\ &\left. \left(\hat{\beta}_{0x}\bm{\hat{x}}+\hat{\beta}_{0y}\bm{\hat{y}}+\hat{\beta}_{0z}\bm{\hat{z}}\right)D_s  
+  \left(E_{0y}\hat{\phi_y}^\prime + E_{0z}\hat{\phi}_z^\prime \right) \right. \\ &\left. \left(0\bm{\hat{x}}+\hat{\bm{\phi}_y}\bm{\hat{y}}+\hat{\bm{\phi}_z}\bm{\hat{z}} \right)D_h\right]
\frac{\mathrm{e}^{-j k (|\bm{s^{'}}|+|\bm{s}|)} }{|\bm{s^{'}}|\sqrt{|\bm{s}|}}. 
\end{split}
\label{eq_electric_field_building_edge}
\end{equation}

\end{corollary}

\begin{IEEEproof}
    The required orthogonal vectors for the incident and diffracted electric field for the upper and lower edge diffraction MPCs are presented in Table \ref{Table_edge_coord_vectors_edges}. Note that the x-component of the vector $\bm{\hat{\phi^{\prime}}}$, $\hat{\phi_x}^\prime=0$ and the x-component of the vector $\bm{\hat{\phi}}$, $\hat{\phi_x}=0$ for both edges. Hence, the diffraction electric field for each edge can be obtained using definition \ref{definition_electric_field_half_plane} and the vectors for each edge. 
\end{IEEEproof}


\subsection{Relative power of the two diffraction MPCs}
\label{sec_rel_power_diffraction_mpc}
\begin{figure}[ht]
    \centering
    \includegraphics[trim=3cm 8cm  4cm 10.0cm,width=0.9\linewidth]{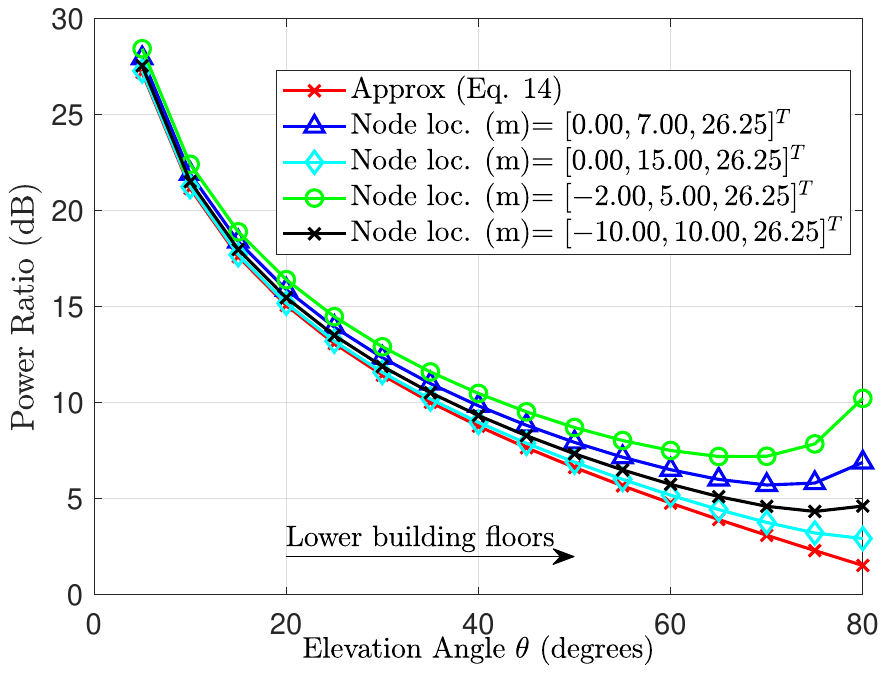}
  \caption{Ratio of Power $\frac{P^u}{P_l}$ in the diffraction MPC for $\theta \in [0,90]$ for various node locations without Assumption \ref{assumption_distance_node_window} compared with the approximation in eq. \eqref{eq_power_ratio} with Assumption \ref{assumption_distance_node_window}. The elevation angle decreases as we go towards higher floors above the UAV anchor.}
  \label{fig_power_ratio}
\end{figure}

In this section, we examine the effect of controlling the polarization of the transmit side electric field on the relative power of the two diffraction paths. First, we derive expressions for the soft and hard diffraction coefficients for the two diffraction MPCs as a function of the elevation angle $\theta$ as in Fig.\,\ref{fig_simplified_building_model}.
From the definition of the diffraction coefficients in eq. \eqref{eq_diffraction_coeff} we first need to obtain the angles $\psi^\prime$ and $\psi$ using eq. \eqref{eq_psi_prime_psi}. We derive $\bm{s_t^\prime}$ and $\bm{s_t}$ for both edges using eq. \eqref{eq_s_t_prime_and_s_t} and the values in Table \ref{Table_edge_incident_diffracted_ray}, Table \ref{Table_edge_coord_vectors_edges}. 
\begin{assumption}
We assume
\label{assumption_distance_node_window}
\begin{itemize}
    \item The window dimension `$w$' is much smaller than the node's distance to the window `$y_n$'.
    \item Define angle $\theta\approx\cos^{-1}{\left(\frac{z_e^l-z_a}{\sqrt{(y_a)^2+(z_e^l-z_a)^2}}\right)} \approx \cos^{-1}{\left(\frac{z_e^u-z_a}{\sqrt{(y_a)^2+(z_e^u-z_a)^2}}\right)}$ as the elevation angle of the UAV-anchor from the perspective of the building floor. 
\end{itemize} 
\end{assumption}
Note, the above assumption is only used to derive insights and is relaxed in the numerical results in Fig. \ref{fig_power_ratio}.
\begin{proposition}
\label{prop_diffraction_mpc_aoa_90}
 Applying the first statement in Assumption \ref{assumption_distance_node_window} to our simplified building model leads to the following implications.
 \begin{itemize}
     \item The diffraction MPCs arrive at the node location approximately parallel to the floor. In other words, the elevation AoA for the diffraction MPCs at the receiver will be approximately \footnote{The AoA is defined as the vertical angle between the MPC propagation direction and the z-axis.Our observation about the AoA of the diffraction MPCs arriving parallel to the floor also matches the observation using Raytracing software in Appendix \ref{section_raytracing_appendix}.}$90^\circ$. This result is obtained rigorously by substituting statement 1 of Assumption \ref{assumption_distance_node_window} into the expression for the angle $\psi$ presented in Table \ref{table_s_t_psi} to obtain $\psi \approx \frac{3\pi}{2}$. The angle $\psi$ is shown in Fig.\,\ref{fig_GTD_half_plane}.
     \item The path lengths of the two diffraction MPCs generated by the upper and lower window edges are approximately the same. This is obtained by applying the assumption to the IPL and OPL defined in Corollary\ref{corollary_diffraction_path_length_building} for both upper and lower edge diffraction MPCs.
 \end{itemize}
 \end{proposition}

\begin{table*}[t]
\caption{$\bm{s^\prime_t}$, $\bm{s_t}$, $\hat{\psi}^\prime$, $\hat{\psi}$ for upper and lower edges.}
\label{table_s_t_psi}
\centering
\begin{tabular}{|c|c|c|}
\hline
 & Lower edge & Upper edge  \\

\hline
$\bm{s^\prime_t}$ & $\frac{-y_a\bm{\hat{y}}+(z_e^l-z_a)\bm{\hat{z}}}{\sqrt{(y_a)^2+(z_e^l-z_a)^2}}$  & $\frac{-y_a\bm{\hat{y}}+(z_e^u-z_a)\bm{\hat{z}}}{\sqrt{(y_a)^2+(z_e^u-z_a)^2}}$   \\
\hline
$\bm{s_t}$ & $\frac{y_n\bm{\hat{y}}+(0.5w)\bm{\hat{z}}}{\sqrt{(y_n)^2+(0.5w)^2}}$ & $\frac{y_n\bm{\hat{y}}-(0.5w)\bm{\hat{z}}}{\sqrt{(y_n)^2+(0.5w)^2}} $   \\
\hline
$\psi^\prime$ & $\arccos{\left(\frac{(z_e^l-z_a)}{\sqrt{(y_a)^2+(z_e^l-z_a)^2}}\right)} \approx \theta $& $\pi-\arccos{\left(\frac{(z_e^u-z_a)}{\sqrt{(y_a)^2+(z_e^u-z_a)^2}}\right)} \approx \pi-\theta$    \\
\hline
$\psi$  & $\pi + \arccos{\left(\frac{(0.5w)}{\sqrt{(y_n)^2+(0.5w)^2}}\right)} \approx \frac{3\pi}{2} $ & $2\pi-\arccos{\left(\frac{(-0.5w)}{\sqrt{(y_n)^2+(0.5w)^2}}\right)\approx \frac{3\pi}{2}}$  \\
\hline
\end{tabular}
\end{table*}

The second statement in Assumption \ref{assumption_distance_node_window} leads to relating angles $\psi^\prime$ for both the diffraction paths with the UAV anchor elevation angle $\theta$ shown in Fig.\ref{fig_simplified_building_model}. The expressions for $\psi^{\prime}$ for both diffraction MPCs are presented in Table \ref{table_s_t_psi}. Observe from Fig.\,\ref{fig_GTD_half_plane}, for each diffraction MPC, $\psi^{\prime}$ is the angle the incident diffraction ray makes with the half-plane, and this suggests that the electric field strength depends on the UAV anchor elevation angle. To investigate the dependence of the electric field strength on the elevation angle $\theta$, we derive approximate expressions for the diffraction coefficients.

\par
\begin{lemma}
\label{lemma_approximated_soft_diffraction_coefficients}
The soft ($D_s$) and hard ($D_h$) diffraction coefficients for the two diffraction MPCs are presented in Table \ref{table_diffraction_coeff}. Note that the superscripts `u' and `l' refer to the upper and lower diffraction MPCs, respectively. 

\begin{table}[htbp]
\caption{$D_{s,h}$ for upper and lower diffraction MPCs.}
\centering
\begin{tabular} {|p{0.22\linewidth}|c|c|} 
\hline
 & Lower edge & Upper edge  \\
\hline
{\footnotesize Soft Diff. Coeff.} & $D_s^l \approx A\left(\frac{-2\sqrt{1-\cos{\theta}}}{\cos{\theta}}\right)$  & $D_s^u \approx A\left(\frac{2\sqrt{1+\cos{\theta}}}{\cos{\theta}}\right)$   \\
\hline
{\footnotesize Hard Diff. Coeff.} &  $D_h^l \approx A\left(\frac{2\sqrt{1+\cos{\theta}}}{\cos{\theta}}\right)$ & $D_h^u \approx A\left(\frac{-2\sqrt{1-\cos{\theta}}}{\cos{\theta}}\right)$   \\
\hline
\end{tabular}
\label{table_diffraction_coeff}
\end{table}

\end{lemma}
\begin{IEEEproof}
Please refer to Appendix \ref{section_appendix_diffraction_coeff_derivation}.
\end{IEEEproof}
To compare the electric field strength of the two diffraction MPCs, we evaluate the ratio of the power contained in each diffraction MPC and consequently show that the ratio is a function of the elevation angle $\theta$. 
\begin{theorem}
\label{theorem_horizontal_polarization}
If at the transmitter side, the electric field is polarized along the x-axis, i.e., parallel to the diffracting edges, then the upper diffraction MPC dominates over the lower diffraction MPC in terms of power. We can express the ratio of power in the two diffraction MPC due to the upper and lower edges purely as a function of the angle $\theta$ as
\begin{align}
\begin{split}
\frac{P^u}{P^l} = \frac{|\bm{E}_{d}^u|^2}{|\bm{E}_{d}^l|^2} \approx \frac{1+\cos{\theta}}{1-\cos{\theta}}.\\
\label{eq_power_ratio}
\end{split}
\end{align}
\end{theorem}

\begin{IEEEproof}
Refer to Appendix \ref{appendix_power_ratio_expression_proof}.
\end{IEEEproof}

In Fig.\,\ref{fig_power_ratio}, we plot the approximate power ratio expression of the two diffraction MPCs in eq. \eqref{eq_power_ratio} as a function of the elevation angle $\theta$ to the anchor. This is compared with the exact power ratio which is calculated without Assumption \ref{assumption_distance_node_window} and depends on the node position within the building and shows good agreement. Note, $\theta=90$\textdegree{} represents the building floor at the same height as the UAV-anchor. Now as we go towards higher floors, $\theta$ decreases and for $\theta<60$\textdegree{}, i.e., for building floors located above the height of the UAV-anchor, the diffraction MPC power ratio is at least 5dB. Therefore, we can conclude that for the locations of the nodes on the upper floors of the building, the diffraction MPC of the upper edge dominates the diffraction MPC of the lower edge, leading to a single {\em dominating diffraction MPC}.

\section{Indoor Positioning}
 Our analysis in Section \ref{fig_simplified_building_model} shows the diffraction MPCs generated by the edges of the windows hold the key to solving the problem of NLoS bias, however these are not the only MPCs generated by interactions with the windows. For the positioning system in Fig.\,\ref{fig_positioning_system}, the impulse response of the multipath wireless channel \cite{heath2016overview} between the UAV anchor and a receiver located within the building can be expressed as 
\begin{equation}
\label{eq_mp_impulse_response}
h(t) = \sum_{k=0}^{K-1}h_k \cdot a_{rx,k}(\theta_{rx,k})\cdot a_{tx,k}(\theta_{tx,k})\cdot \delta(t-\tau_k).
\end{equation}
Here, $a_{rx,k}(\theta_{rx,k})$ and $a_{tx,k}(\theta_{tx,k})$ are the respective gains of the receiver and transmitter antenna for the $k^{th}$ MPC formed between the transmitter and the receiver. Each MPC has four properties: (a) complex gain: $h_k$, (b) Angle-of-Arrival (AoA): $\theta_{rx,k}$, (c) Angle-of-Departure(AoD): $\theta_{tx,k}$ and (d) the propagation delay: $\tau_k$. We assume that the transmitter and receiver antenna gains only vary across elevation, thus the path gain is only affected by the elevation AoA and AoD of the MPCs. Note that $\theta$ in Assumption $\ref{assumption_distance_node_window}$ is slightly different from AoD $\theta_{tx,k}$. Both AoA $\theta_{rx,k}$ and AoD $\theta_{tx,k}$ are the angles the MPCs make with the z-axis. The four parameters for each MPC are ultimately affected by objects in the environment and the frequency of operation through the three main mechanisms: (a) reflection from surfaces, (b) diffraction from edges, and (c) transmission through various building materials. Both the building materials and the geometry of the walls can attenuate the MPCs by changing the number of significant reflections, diffractions, and transmissions each path can undergo before it reaches the receiver. 
We conducted a statistical study on the MPCs generated for an O2I signal propagation scenario relevant to our positioning system. The key results used for further analysis are shown below, with the details of the study presented in Appendix \ref{section_raytracing_appendix}.
\begin{itemize}
    \item The diffraction MPCs between the outdoor transmitter and the indoor receiver are generated by a single interaction of the propagating signal with the upper or lower edge of the windows located on the same floor of the building as the receiver. For an \footnote{The first arriving path will be one of the diffraction MPC assuming the direct path is sufficiently attenuated, however the direct path attenuation is ultimately a function of the operating frequency and building material and possibly holds for mmWave frequencies. The operating frequency $28$GHz assumption is based on the measurement campaign in \cite{bas2019outdoor}.  }operating frequency of 28 GHz and brick-walled buildings, these diffraction MPCs are the first arriving paths at most of the indoor receiver locations.     
\end{itemize}

\subsection{ToF based range measurements of the first arriving path}\label{section_tof_range_measurements}
To achieve our ultimate goal of 3D positioning, we are interested in estimating the path delay associated with the diffraction MPCs. Based on the previous section, this is ultimately equivalent to estimating the propagation delay of the first arriving path between the outside anchor and the indoor node. The signal's propagation delay along the first arriving path will correspond to one of the two diffraction paths and can be estimated using the received signal measurements. Assume that the anchor and node are perfectly time synchronized\footnote{This is not a strict requirement. We could also use systems based on round trip time (RTT) as specified in 3GPP and WiFi standards}. Now, we transmit a pulse of power $E_p$ and pulse shape $p(t)$ of duration $T_p$ from the transmitter.
Convolving the transmit pulse with the channel defined in eq. \eqref{eq_mp_impulse_response} we obtain the received signal in the presence of Gaussian noise as 
\begin{align}
\begin{split}
r(t) &= \sqrt{E_p} \sum_{k=0}^{K-1}h_k  a_{rx,k}(\theta_{rx,k}) a_{tx,k}(\theta_{tx,k}) p(t-\tau_k)+ n(t) \\
&= \sum_{k=0}^{K-1} \gamma_k(\theta_{tx,k},\theta_{rx,k}) p(t-\tau_k) + n(t).
\end{split}
\end{align}
Here, $\gamma_k(\theta_{tx,k},\theta_{rx,k})$ is the complex gain for the received signal due to the $k^{th}$ MPC and includes the effects of the transmit and receive side antennas, transmitted signal power and also the path gain $h_k$. For further analysis, the receiver noise $n(t)$ is assumed to be Gaussian. Of the $K$ MPCs in the received signal, let the indices $k=0$ and $k=1$ correspond to the two diffraction MPCs. Therefore, the remaining $K$-$2$ MPCs consist of those MPCs that contain more than one interaction with the environment. If the electric field for the transmit signal is parallel to the horizontal diffracting window edges, Theorem \ref{theorem_horizontal_polarization} applies. Therefore, it leads to the upper edge diffraction MPC dominating over the lower edge diffraction MPC. In a system this could be achieved by mounting the linear polarized antennas horizontally, instead of the conventional vertically mounted implementation. Thus, the received signal can be written as 


\begin{align}
\begin{split}
&r(t) \approx \underbrace{ \gamma_0(\theta_{tx,0},\theta_{rx,0}) p(t-\tau_0)}_{\text{dominating diffraction MPC}} \\
&\hspace{5em}+ \underbrace{\sum_{k=2}^{K-1} \gamma_k(\theta_{tx,k},\theta_{rx,k}) p(t-\tau_k)}_{\text{MPC-4}}+ n(t)\\
\end{split}
\end{align}


Next, we assume resolvability between the MPCs i.e., $|\tau_i-\tau_j| > T_p, \;\; \forall \;\; i\neq j $. It can be shown with this assumption, that the CRLB on the RMSE for estimating the propagation delay of the first arriving path $\tau_0$ is given by \cite{falsi2006time,van2004detection}

\begin{equation}
\label{eq_ranging_CRB}
\begin{split}
    \text{CRLB}(\tau_0) = \frac{1}{\sqrt{8\pi^2\beta^2\text{SNR}}} 
    = \sqrt{\frac{N_0}{8\pi^2\beta^2(\gamma_0(\theta_{rx,0},\theta_{tx,0}))^2}}.
\end{split}
\end{equation}
Here, $\gamma_0(\theta_{rx,0},\theta_{tx,0})$ is the amplitude corresponding to the first arriving path and includes the transmit and receive side antenna gains, SNR is receive side signal-to-noise ratio (linear), and $\beta$ is the bandwidth of the transmitted signal (Hz). Since this first arriving path corresponds to the {\em dominating diffraction path}, we have $\tau_0 = \frac{p}{c}$.
Here, $\tau_0$ is the propagation delay of the first arriving path, and the path length of the dominating diffraction path $p$ is given by eq. \eqref{eq_path_length_building} and $c$ is the speed of light. Drawing insights from the raytracing analysis in Appendix \ref{section_raytracing_appendix} and eq. \eqref{eq_ranging_CRB}, we can improve the estimate of $p(\bm{\alpha})$ by maximizing the receiver side antenna gain $\alpha_{rx,k}(\theta_{rx,k})$ for AoA $\theta_{rx,k} = 90$\textdegree{}. In other words, we could improve the ranging performance if we orient the receiver side antenna such that the antenna gain is maximum towards the windows. Further improvement to the ranging estimate can be made by scanning the transmit side beam to maximize the receiver side power of the dominant diffracting MPC.

\subsection{Fisher Information analysis of the 3D position estimate}\label{section_FIM_derivation}
\begin{figure}[ht]
    \centering
    \includegraphics[width=0.8\linewidth]{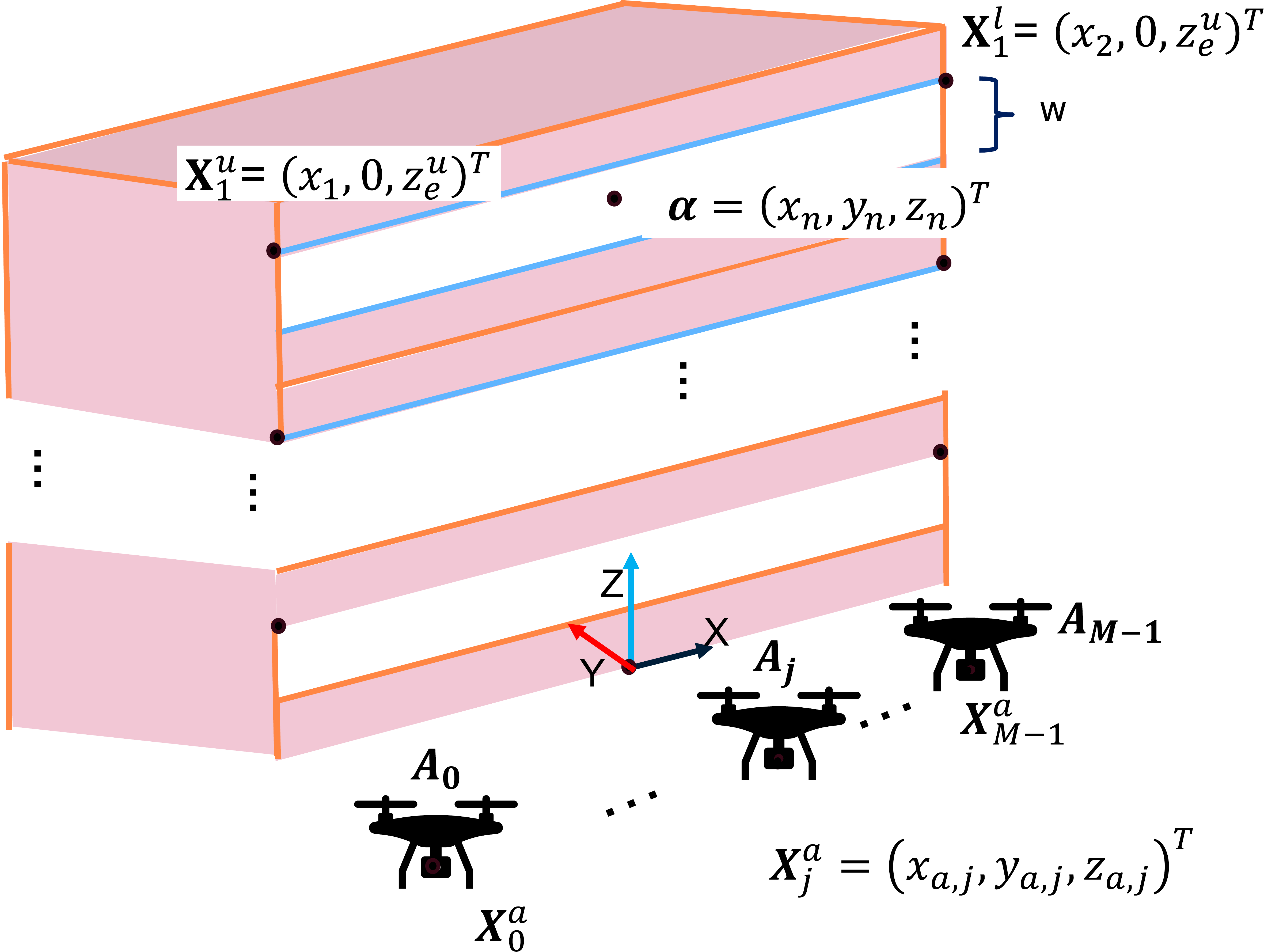}
  \caption{3D Positioning model for FIM analysis}
\label{fig_positioning_model}
\end{figure}

In this section, we examine the conditions under which we can perform 3D NLoS localization using the ToF-based ranging estimates corresponding to the {\em dominating diffraction} MPC using Fisher Information analysis and then calculate the lower bound to the 3D position error. In Fig.\ref{fig_positioning_model}, we have `$M$' UAV anchors in front of the building. The node is located on an arbitary building floor which has two diffracting edges separated by a vertical distance $w$. We assume that the origin is located in the middle bottom of the front wall of the building at the intersection of the front wall with the ground, the x axis oriented along the diffracting edges and the z-axis pointed along the vertical direction as in Fig.\ref{fig_positioning_model}.

Each anchor indexed by `$j$' is placed at known coordinates $\bm{X}^a_j = [x_{a,j},y_{a,j},z_{a,j}]^T$ relative to the origin, while the node is located on an arbitrary floor of the building at $\bm{\alpha}=[x_n,y_n,z_{n}]^T$. 
Assuming that the anchors transmit x-polarized signals, each ToF-based range measurement corresponds to the path length of the respective {\em dominant diffraction MPC} between each anchor $j$ and the node. With a slight change in notation, we denote the path length of the respective {\em dominating diffraction MPC} as $p_j(\bm{\alpha})$ obtained using eq. \eqref{eq_path_length_building}, with the subscript `$j$' representing the $j^{\text{th}}$ anchor. Consequently, we obtain $M$ noisy range measurements from which we need to estimate the 3D position of the node. Assuming the 3D position of the node $\bm{\alpha}$ as the vector of the three parameters to be estimated, its corresponding Fisher Information Matrix (FIM) $\bm{J}_{\bm{\alpha}} \in \mathbb{R}^{3 \times 3}$ \cite{kay1993fundamentals}, \cite{zekavat2011handbook} and can be expressed as

\begin{equation}
\label{eq_FIM}
\bm{J}_{\bm{\alpha}} = c^2\bm{J}^T \bm{J}_{\tau_0} \bm{J}.
\end{equation}
Since we assume the noise in our range measurements is independent across the $M$ anchors, $\bm{J}_{\tau_0} = diag(\lambda_0, \cdots,\lambda_{M-1})$ is a diagonal matrix where $\lambda_j = 8\pi^2\beta^2SNR_j$ represents the inverse of the CRLB of the path delay estimate of the first arriving path between the $j^{th}$ anchor and the node. The Jacobian $\bm{J} \in \mathbb{R}^{3 \times M}$ is the transform from the ranging measurement space $\in \mathbb{R}^M$ to the unknown parameter to be estimated space $\in \mathbb{R}^3$ and is given by   

\begin{equation}
\label{eq_Jacobian}
\bm{J} = 
\begin{bmatrix}
\frac{\partial{p_{0}(\bm{\alpha})}}{\partial x_n} & \cdots &  \frac{\partial{p_{M-1}(\bm{\alpha})}}{\partial x_n}\\
\frac{\partial{p_{0}(\bm{\alpha})}}{\partial y_n} &\cdots &  \frac{\partial{p_{M-1}(\bm{\alpha})}}{\partial y_{n}}\\
\frac{\partial{p_{0}(\bm{\alpha})}}{\partial z_n} &\cdots &  \frac{\partial{p_{M-1}(\bm{\alpha})}}{\partial z_{n}}
\end{bmatrix}.
\end{equation}

The node's unknown 3D coordinates $\bm{\alpha}$ are estimatable if and only if the FIM $\bm{J}_{\bm{\alpha}}$ is invertible \cite{don_ris_near_far_field}, therefore in the further analysis we study the conditions for its invertibility.

\begin{theorem}
\label{theorem_3Dlocalization}
The FIM $\bm{J}_{\bm{\alpha}} \in \mathbb{R}^{3\times3}$ defined in eq. \eqref{eq_FIM} is invertible if and only if $rank(\bm{J})=3$. Here, the Jacobian Matrix $\bm{J} \in \mathbb{R}^{M\times 3}$ is defined in eq. \eqref{eq_Jacobian} from the ToF ranging measurement space $\mathbb{R}^{M}$ to the 3D coordinate space  $\bm{\alpha} \in \mathbb{R}^{3}$. 
\end{theorem}

\begin{IEEEproof}
From eq. \eqref{eq_FIM}, observe that $\bm{J}_{\tau_0} \in \mathbb{R}^{M \times M}$ is a diagonal matrix with non zero diagonal entries since it represents the FIM of the independent ranging measurements from `$M$' anchors. Hence, we can write  
$$\bm{J}_{\bm{\alpha}} = c^2\bm{J}\bm{J}_{\tau_0} \bm{J}^T = c^2\bm{J}\sqrt{\bm{J}_{\tau_0}}\left(\sqrt{\bm{J}_{\tau_0}} \bm{J} \right)^T.
 $$
 $\sqrt{\bm{A}}$ is the elementwise square root operation of a matrix $\bm{A}$.
 We also know that multiplying a matrix with a non-zero diagonal matrix does not change its rank. Furthermore, by using $rank(\bm{A}\bm{A}^T)=rank(\bm{A})$  \cite{strang2020linear}, we obtain $rank(\bm{J}_{\bm{\alpha}})=rank(\bm{J})$. Now, $\bm{J}_{\bm{\alpha}}$ is invertible if and only if $rank(\bm{J}_{\bm{\alpha}})=3$. Hence, we obtain the desired result.   
\end{IEEEproof}

\begin{corollary}
\label{corollary_conditions_for_positioning}
    To obtain the 3D position estimate for the node at $\bm{\alpha}=[x_n,y_n,z_{n}]^T$ for the scenario shown in Fig.\,\ref{fig_positioning_model} the following conditions must be met: 
\begin{itemize}
\item It is necessary but not sufficient to have at least three anchors. This is because, for $rank(\bm{J})=3$, it is necessary but not sufficient to have $M \geq 3$. 
\item The position of the anchors $\bm{X}^a_j=[x_{a,j}, y_{a,j}, z_{a,j}]^T \; \forall j \in[0,M-1]$ is known in local building coordinates.
\item The x-coordinates of the endpoints of the upper edges of the diffracting edges i.e., $x_1$ in $\bm{X}_1^{U}, x_2$ in $\bm{X}_2^{U}$ and the window dimension $w$ need to be known {\em a priori}.
\item The $Z$-coordinate $z_{e}^u$ of the diffracting edges on the node building floor is not required to be known.         
\end{itemize}
\end{corollary}
This result demonstrates that we no longer require detailed position knowledge about the diffraction edges which is an improvement over the conference version of this paper \cite{duggal20243d} that relied on the path model in Lemma \ref{lemma_half_plane_path_length}. Essentially by incorporating the fact that the diffraction occurs on the same floor as the UE and that the vertical coordinate of the diffraction point is very close to that of the node, we are able to reduce one degree of freedom from the path model in Lemma \ref{lemma_half_plane_path_length} to Corollary \ref{corollary_diffraction_path_length_building}.
\par
To illustrate that having $M=3$ anchors is not a sufficient condition for 3D positioning, consider a scenario where we have three anchors such that their x-coordinate $x_{a,j}=0\;\; \forall\;\; j \in [0,2]$. Observe that as the node position is varied inside the building such that it approaches the plane $X=0$, i.e., $x_n\to 0$, for each anchor, we have the x-coordinate of the corresponding diffraction point $q_x \to 0$. Notice in the first row of the Jacobian $\bm{J}$, defined in eq. \eqref{eq_jacobian_elements}, the term $\frac{\partial p_j{(\bm{\alpha}})}{\partial x_n} \to 0\;\;\forall j\in[0,2]$. Therefore, for $x_n=0$, $rank(\bm{J}) \neq 3$, and by using Theorem \ref{theorem_3Dlocalization}, we cannot estimate the 3D position of the node for this scenario. 
\par
The diffracting edges are located parallel to the X-axis at $Y=0$. Observe the x-coordinates of the end points of the diffracting edges correspond to the width of the building which is required to be known {\em a priori} whereas the height of the diffracting edges is not required to be known. This is because the node and the edge from which diffraction takes place are located on the same building floor. Therefore, by estimating the node location leads to an estimate of the height of the edge since the diffracting edge is located at a distance $0.5w$ above the node.  Note Corollary \ref{corollary_conditions_for_positioning} leads to the following observation - we need position knowledge of the anchors in local building coordinates. These local coordinates are defined with respect to the origin located on the diffracting edges as shown in Fig.\,\ref{fig_positioning_model}. 
\par
One of the ways to realize a real system, from Corollary \ref{corollary_conditions_for_positioning}, is to equip the UAV anchors with GNSS systems, which will provide position knowledge in a global coordinate frame. The translation of these global anchor coordinates to the local building coordinates can be obtained by using map information about the building where the edges of the building can be positioned in the global frame. Alternatively, in areas where GNSS systems may not work accurately, we could use different kinds of sensors to place the anchors in the local building frame. For example, the distance of the anchor to the building $y_{a,j}$ can be obtained using lidar or a laser range finder to the building wall, the elevation of the UAV anchor from the ground $z_{a,j}$ can be obtained using barometers and $x_{a,j}$ can be obtained with inter-anchor ToF ranging measurements.   
\section{Simulation Results}
In this section, we first describe a simulation strategy to generate noisy range estimates from raytracing simulation data. Using these range measurements, we then propose a non-linear least squares algorithm called `D-NLS' for 3D positioning based on our diffraction path model. For comparison, we implement two alternative approaches: (a) Linear Least Squares (LLS), which assumes that all range measurements are LoS, and (b) the Iterative Parallel Projection Algorithm (IPPA) for positioning in NLoS scenarios \cite{jia2010set,jiacol}. The IPPA is a set-theoretic method that mitigates the impact of NLoS bias by utilizing information about whether a range measurement is LoS or NLoS. Los/NLos identification can be achieved by using techniques in \cite{venkatesh2007non}. 
\par
From our raytracing simulation in Appendix \ref{section_raytracing_appendix}, using a thresholding logic we pick the first arriving path between each candidate Rx location and each UAV anchor. Observe from Table \ref{table_window_MPC} the first arriving path across receivers are majorly the diffraction paths i.e. MPCs present in MPC-3. These should have a true path length derived in Corollary \ref{corollary_diffraction_path_length_building}. Our measurement model for the range measurements at each Rx location $\bm{\alpha}$ is
\begin{equation}
    \bm{r} = \mathbf{p(\bm{\alpha}) + \mathbf{n}}.
    \label{eq_measurement_model}
\end{equation}
Here, $\bm{r}=[r_0,\cdots,r_j,\cdots,r_{M-1}]^T \in \mathbb{R}^M$ is the vector of the noisy range measurements from $M=4$ anchors, $\mathbf{p(\bm{\alpha})}=[p_0(\bm{\alpha}),\cdots, p_{j}(\bm{\alpha}), \cdots,p_{M-1}(\bm{\alpha})]^T$, where $p_j(\bm{\alpha})$ is the diffraction path length between the $j^{\text{th}}$ anchor and node obtained using Corollary \ref{corollary_diffraction_path_length_building}. Now the uncertainty in range measurements is limited by both SNR and signal Bandwidth. Hence from Section \ref{section_tof_range_measurements}, the noise vector is simulated as $\bm{n}=[n_0,\cdots,n_{M-1}]$, where $n_{j} \sim \mathcal{N}(0,\sigma_j^2)$, $\sigma_j= \frac{c}{\sqrt{8\pi^2\beta^2\text{SNR}_{j}}}$ is the uncertainty in the range measurements for $j^{\text{th}}$ anchor.

\begin{table}[ht]
\caption{Simulation Parameters.}
\label{table_simulation_params}
\centering
\small
\begin{tabular}{|c|c|}
\hline
Description & Symbol \& Value   \\
\hline
Anchor $1$ pos. & $\bm{X}_0^a=[-15m, -20m, 1m]^T$ \\
\hline
Anchor $2$ pos.& $\bm{X}_1^a=[0m, 15m, 0m]^T$ \\
\hline
Anchor $3$ pos.& $\bm{X}_2^a=[0m, -7m, 4m]^T$ 
\\
\hline
Anchor $4$ pos.& $\bm{X}_3^a=[15m, -20m, 1m]^T$ 
\\
\hline
Signal Bandwidth & $\beta=400$ MHz \\
\hline
Noise Floor &  $N_0$=-87 dBm \\
\hline
Rx processing gain & 20dB \\
\hline
\end{tabular}
\end{table}

\subsection{3D positioning algorithm based on the diffraction path model}
\label{section_non_linear_least_squares}
From the measurement model in eq. \eqref{eq_measurement_model}, we have the noisy range measurements $\bm{r}$ and the UAV anchor positions from which we need to obtain a position estimate $\bm{\hat{\alpha}}$. This can be achieved by minimizing the least squared error between the range measurements and the diffraction path model as $$ \bm{\hat{\alpha}} = \underset{\bm{\alpha}}{\mathrm{\text{arg min}}} \sum_{j=1}^M| r_{j}-{p_j}(\bm{\alpha})| ^2. $$   
Since the diffraction path length model $p_j(\bm{\alpha})$ is a non-linear function we utilize a Gauss-Newton approach \cite{kay1993fundamentals} to derive an iterative estimator referred to as \textbf{D-NLS (Case-1)}. Using subscript $m$ to denote the iteration index the iterative update steps are
\begin{equation}
    \bm{\hat{\alpha}}_{m+1} = \bm{\hat{\alpha}}_{m} + (\bm{J}_{m}^T\bm{J}_m)^{-1}\bm{J}_m^T(\bm{r}-\bm{p}(\bm{\hat{\alpha}}_m)). 
\end{equation}
Here $\bm{J}_m$ is the Jacobian matrix defined in eq. \eqref{eq_Jacobian} for the $m^{\text{th}}$ iteration. In the analysis until now, we had assumed that the node is located at the midpoint of the building floor as proposed in Assumption \ref{assumption1_diffraction_same_floor}. However, this may not necessarily hold true as the indoor user might lay prone on the floor or the receiver may be mounted at a different height. We account for this by assuming that there is a distance offset $\Delta$ between the node's z-coordinate and the upper diffraction edge, i.e. 
\begin{equation}
\label{eq_assumption1_vertical_offset}
z_n = z_e^u-0.5w+\Delta.
\end{equation}
Clearly, $\Delta$ is upper bounded by half the height of the building floor, and by including this offset in our path model in our least squares formulation, we obtain positioning performance which includes a slight model mismatch. Note, $\Delta$ is not estimated as part of the positioning algorithm. It merely represents the amount of model mismatch between our assumption that the z-coordinate of the node is at the mid-point of the building floor and reality. Thus we compare performance with and without Assumption \ref{assumption1_diffraction_same_floor} by comparing D-NLS (Case-1) vs D-NLS (Case-2). Note, in D-NLS (Case-2) which includes a model mismatch sets an upper bound to our proposed diffraction-based positioning technique. 

\subsection{Results and discussion}
\begin{figure*}
    \subfigure[3D positioning Performance.]{ 
        \begin{minipage}[t]{0.49\textwidth}
        \centering
        \includegraphics[ width=0.8\linewidth]{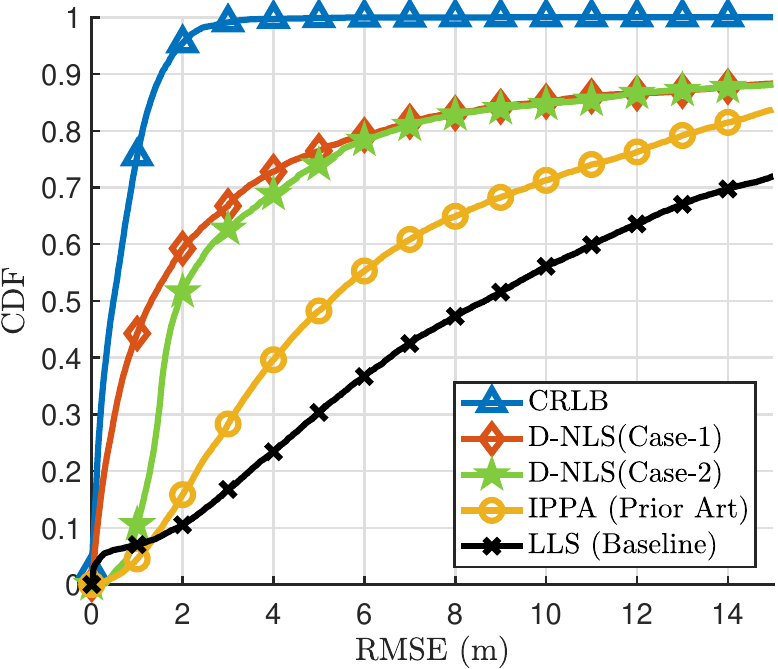}
        \label{fig_3D_positioning}    
        \end{minipage}}
        \subfigure[Z-axis positioning Performance.]{
        \begin{minipage}[t]{0.49\textwidth}
        \centering
        \includegraphics[ width=0.8\linewidth]{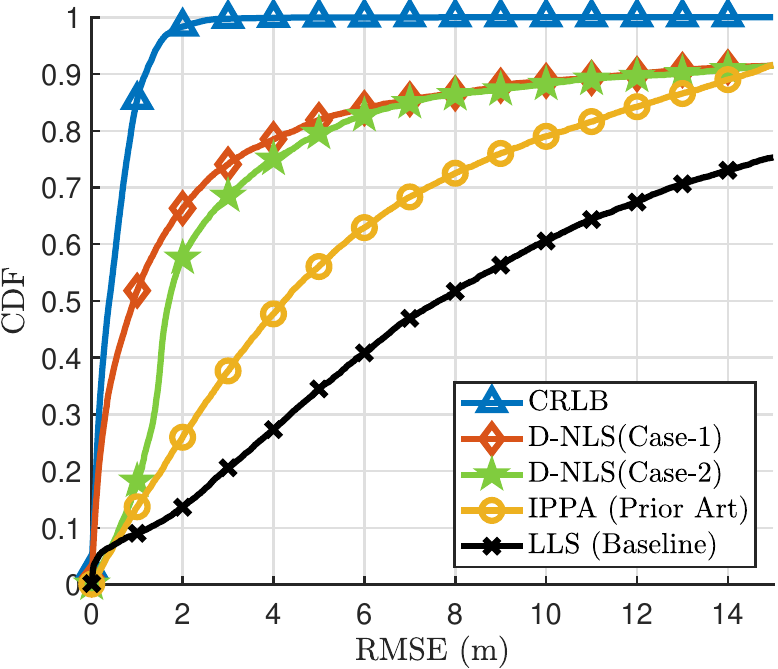}
        \label{fig_results_Z_positioning}  
        \end{minipage}}

        \caption{
We evaluate the positioning performance in NLoS conditions by comparing the lower bound derived using the CRLB for our proposed diffraction-inspired positioning technique (D-NLS), both with and without Assumption \ref{assumption1_diffraction_same_floor} (Case-1 vs. Case-2). Our approach demonstrates improved positioning performance compared to both the prior art in NLoS positioning (IPPA) and the baseline method (LLS) for (a) 3D positioning and (b) z-axis positioning accuracy.} 
        \label{fig_positioning_results}
 \end{figure*}

To obtain the theoretical lower bound to the performance, we use the Position Error Bound (PEB) obtained as $\text{PEB}=\sqrt{\text{Tr}({\bm{J_{\alpha}}}^{-1})}$ where $\text{Tr}{(\bm{A})}$ is the trace of matrix $\bm{A}$ and $\bm{J_{\alpha}}$ is the FIM from Section \ref{section_FIM_derivation}. Similarly, the bound on the z-axis is defined as $\sqrt{{(\bm{J}^{-1}_{\bm{\alpha}})}^{[3,3]}}$, where the notation $\bm{A}^{[i,j]}$ represents the element in the $i^{th}$ row and $j^{th}$ column of the matrix $\bm{A}$. Since we have the ground-truth positions of Rx from the raytracing study, we can calculate the 3D RMSE=$\sqrt{(\bm{\alpha}-\bm{\hat{\alpha}})^T(\bm{\alpha}-\bm{\hat{\alpha}})}$ for each positioning approach. Similarly, we can calculate the RMSE in z-axis estimate. Finally, we present the CDF of the RMSE and PEB in Fig. \ref{fig_positioning_results}. The LLS approach serves as the baseline comparison but performs poorly because it assumes LoS conditions, and due to poor anchor geometry since they are located on one side of the building. IPPA is able to improve the positioning performance but due to its reliance on modeling the path length using Euclidean distance, it is limited in its effectiveness at mitigating the positioning errors caused by the NLoS bias in the range measurements. D-NLS (Case-1), our proposed NLS technique based on the diffraction path model including Assumption \ref{assumption1_diffraction_same_floor}, offers the best algorithmic performance, but does not meet the CRLB. This is because of two main reasons (a) model mismatch (b) the NLS approach is sensitive to initialization values and sometimes converges to local minima \cite{zekavat2011handbook}. The first arriving paths are not always diffraction paths (i.e., not included in MPC-3, as defined in Appendix \ref{section_raytracing_appendix}). This may occur either because these paths do not exist at certain Rx locations or because their SNR is extremely low, resulting in missed detections by our first-arriving path estimator. Therefore, in these cases, we assume that the range measurements are due to diffraction in eq. \eqref{eq_measurement_model} would be incorrect and this would be a case of mismatch of the model. Note that in our simulation, we also simulate relaxing Assumption \ref{assumption1_diffraction_same_floor} in D-NLS(Case-2) , which assumes that the node positions are at the midpoint of the building floor. To account for this, we introduced an offset $\Delta=1.5m$, corresponding to half the height of a building floor, in our diffraction path model according to eq. \eqref{eq_assumption1_vertical_offset}. Although including this offset slightly degrades positioning performance as shown in D-NLS (Case-2), our proposed technique still achieves better positioning accuracy in NLoS scenarios compared to previous methods.

\section{Conclusion}

This paper explores positioning in NLoS scenarios, focusing on diffraction as a signal propagation mechanism. Leveraging the Geometrical Theory of Diffraction (GTD), we propose a novel path model for NLoS scenarios where propagation paths are formed through interactions with edges. Based on this model, we develop a positioning technique for a public safety scenario addressing the need for improved 3D and z-axis position estimates where the O2I signal propagation is primarily due to diffraction from window edges through diffraction MPCs. Theoretical analysis based on GTD presents several insights about these diffraction MPCs, such as the presence of two diffraction MPCs resulting from diffraction from the upper and lower edges of the window located on the node building floor. In addition, these MPCs for diffraction arrive in parallel horizontal plane with AoA $\approx 90$\textdegree{}. Additionally, by changing the transmit side electric field polarization we can control the relative power in the two diffraction MPCs.
In the second part of the paper, we establish the necessary and sufficient conditions for estimating the unknown 3D node position inside the building by using Fisher information analysis, and also derive the CRLB for 3D positioning accuracy for our proposed positioning technique. Further, we derive a non-linear least squares algorithm based on our proposed diffraction path model (D-NLS). We also present a realistic raytracing based study that analyses the different MPCs generated for our positioning scenario using statistical metrics. The generated MPCs are also used to generate numerical positioning results showing an improvement in positioning performance for our proposed positioning technique compared to the prior art.

\appendix
\subsection{Derivation of the path length}\label{derivation_path_length}
Consider Fig.\,\ref{fig_GTD_half_plane}; our goal is to derive the geometrical path length $\bm{X_{a}Q_e\alpha}$. We start by calculating the coordinates of the diffraction point $\bm{Q_e}$. For this, we substitute the incident ray unit vector $\bm{\hat{s^{\prime}}}$, edge unit vector $\bm{\hat{e}}$ and diffracted ray unit vector $\bm{\hat{s}}$ in terms of the various coordinates shown in Fig.\,\ref{fig_GTD_half_plane} in the diffraction law eq. \eqref{eq_law_of_diffraction}. Next, we substitute the parametric coordinates of the diffraction point $\bm{Q_e}$ in terms of the parameter $\lambda$ and the endpoints of the diffracting edge $\bm{X_1}$ and $\bm{X_2}$. After some algebraic manipulations, we obtain a quadratic equation for the parameter $\lambda$. All these steps are shown in eq. \eqref{eq_derivation_path_length}.
\begin{figure*}
\begin{equation}
\label{eq_derivation_path_length}
\begin{split}
  & \left| \frac{(q_x-x_a)\bm{\hat{x}} - y_a \bm{\hat{y}} + (z_e-z_a) \bm{\hat{z}}}{\sqrt{(q_x-x_a)^2+y_a^2+(z_e-z_a)^2}} \cdot \bm{\hat{x}} \right| =    \left| \frac{(x_n-q_x)\bm{\hat{x}} + y_n \bm{\hat{y}} + (z_n-z_e) \bm{\hat{z}}}{\sqrt{(x_n-q_x)^2+y_n^2+(z_n-z_e)^2}} \cdot \bm{\hat{x}} \right| \\
  &\Rightarrow  \left| \frac{(q_x-x_a)}{\sqrt{(q_x-x_a)^2+y_a^2+(z_e-z_a)^2}}  \right| 
  =    \left| \frac{(x_n-q_x) }{\sqrt{(x_n-q_x)^2+y_n^2+(z_n-z_e)^2}}  \right| \\
&\Rightarrow   (q_x-x_a)^2 ((x_n-q_x)^2+y_n^2+(z_n-z_e)^2)  =    (x_n-q_x)^2  ((q_x-x_a)^2+y_a^2+(z_e-z_a)^2) \\
&\Rightarrow (\lambda (x_1-x_2) + x_2-x_a)^2 ((x_n-(\lambda (x_1-x_2) + x_2)))^2+y_n^2+(z_n-z_e)^2)  \\ &\;\;\;\;\;\;\; =    (x_n-(\lambda (x_1-x_2) + x_2))^2  (((\lambda (x_1-x_2) + x_2)-x_a)^2+y_a^2+(z_e-z_a)^2) \\
&\Rightarrow a\lambda^2+b\lambda+c = 0.
\end{split}
\end{equation}
\end{figure*}

\subsection{Derivation Of Soft Diffraction Coefficients For The Simplified Building Model}\label{section_appendix_diffraction_coeff_derivation}
Starting from eq. \eqref{eq_diffraction_coeff}, we use the cosine of a sum of angles identities and then the half angle identities $\sin\frac{\psi}{2} = \sqrt{\frac{1-\cos\psi}{2}}$, $\cos\frac{\psi}{2} = \sqrt{\frac{1+\cos\psi}{2}}$. Using the values of $\psi$ and $\psi^\prime$ from Table \ref{table_s_t_psi} for the two edges, we arrive at the final expressions in Table \ref{table_diffraction_coeff}.
\begin{equation}
\begin{split}
  D_s &= A \left( \frac{1}{\cos\frac{\psi-\psi^\prime}{2}} - \frac{1}{\cos\frac{\psi+\psi^\prime}{2}} \right) \\
  & = A \left(\frac{-2\sin\frac{\psi}{2}\sin\frac{\psi^\prime}{2}}{(\cos\frac{\psi}{2}\cos\frac{\psi^\prime}{2})^2-(\sin\frac{\psi}{2}\sin\frac{\psi^\prime}{2})^2} \right)\\
  &= A \left( \frac{-2\sqrt{1-\cos\psi^\prime}}{\cos\psi^\prime}\right).
\end{split}
\label{eq_soft_diffraction_approx}
\end{equation}

\subsection{Power Ratio Of The Diffraction MPCs} \label{appendix_power_ratio_expression_proof}
Observe that in eq. \eqref{eq_electric_field_building_edge}, if the incident electric field is x-polarized i.e., $E_{0,y}=E_{0,z}=0$ the electric field expression simplifies to
\begin{equation}
\begin{split}
\bm{E}^d & = 
\begin{aligned}
\left(E_{0x}\hat{\beta}_{0x}^\prime D_s\right)\left(\hat{\beta}_{0x}\bm{\hat{x}}+\hat{\beta}_{0y}\bm{\hat{y}}+\hat{\beta}_{0z}\bm{\hat{z}}\right)  \\
\end{aligned}
\frac{\mathrm{e}^{-j k (|\bm{s^{'}}|+|\bm{s}|)} }{|\bm{s^{'}}|\sqrt{|\bm{s}|}}. 
\end{split}
\label{eq_efield_x_polarized}
\end{equation}
Observe, by controlling the polarization, we have eliminated the hard diffraction coefficient from our electric field expression. Now, we calculate the ratio of the squared magnitude of this simplified electric field expression and then substitute the approximated soft diffraction coefficients derived in Lemma \ref{lemma_approximated_soft_diffraction_coefficients} and presented in Table \ref{table_diffraction_coeff}. Note that $|\bm{s^\prime}|$ and $|\bm{s}|$ are the OPL and IPL defined in Corollary \ref{corollary_diffraction_path_length_building} respectively and are of similar length according to Assumption \ref{assumption_distance_node_window}. The power ratio result follows from this.

\subsection{Raytracing Simulation}\label{section_raytracing_appendix}
\begin{figure*}
  \centering
\includegraphics[width=0.8\linewidth]{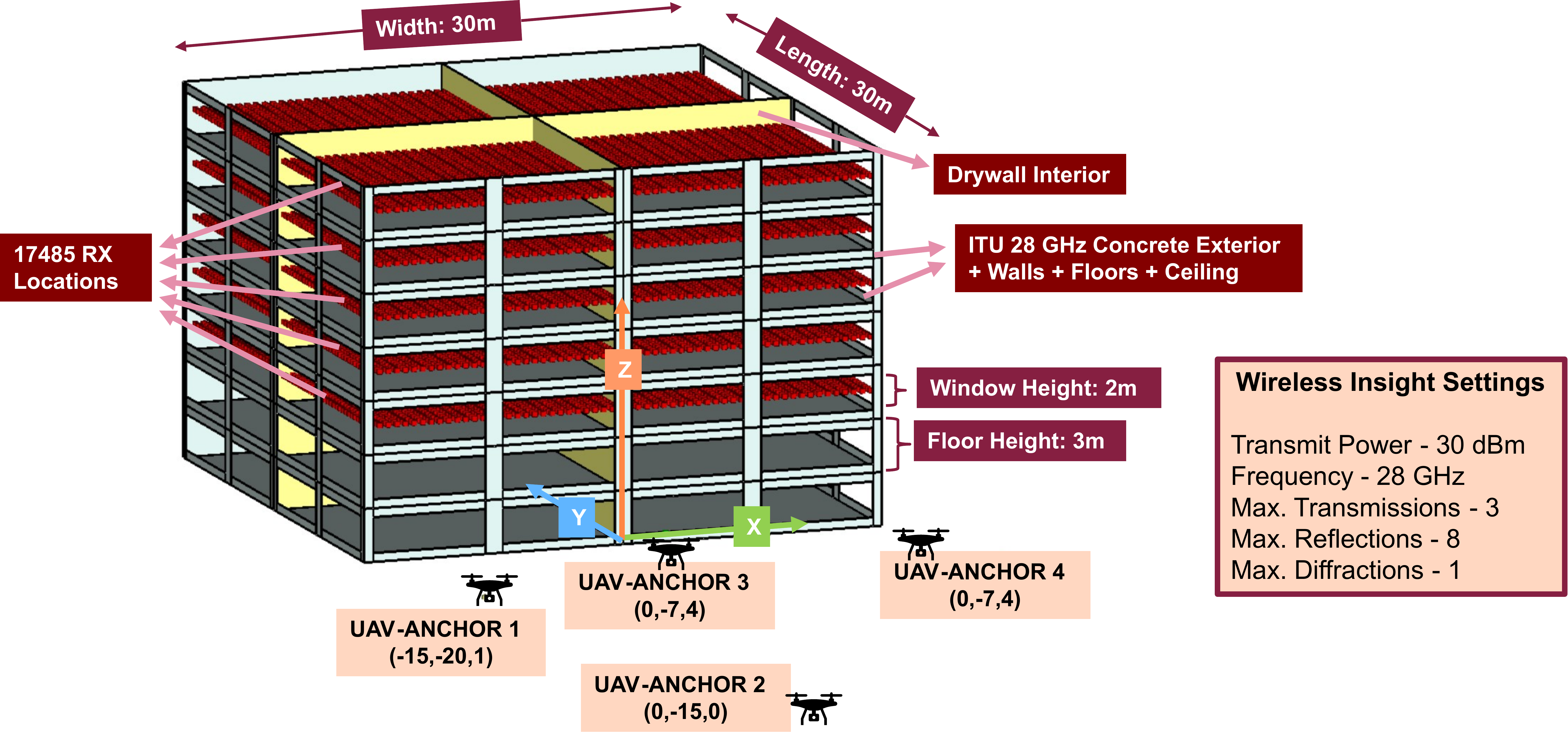}
  \caption{Raytracing setup using Remcom's Wireless Insight \cite{wirelessinsight}. We create a building with floors, ceilings and exterior made out of ITU 28 GHz Concrete and Drywall partitions in the interiors. For numerical results we consider 17,485 candidate receiver locations located across five building floors which are NLoS to all four UAV-anchors situated outside the building.}
  \label{fig:raytracing_wireless_insight}
\end{figure*}

 We set up a model of a building in a realistic wireless raytracing software \cite{wirelessinsight} as shown in Fig.\, \ref{fig:raytracing_wireless_insight}. Further, we consider candidate receiver (Rx) locations spread across five building floors, where each floor is located above the UAV anchor which is equipped with a radio transmitter (Tx). Note, for majority of the candidate Rx locations across floors this would represent NLoS conditions. According to the observation in {\em Bas et al.} \cite{bas2019outdoor}, for each NLoS Rx location, we assume that the direct path is sufficiently attenuated and that all MPCs are generated due to interactions with the windows. We also assume that the windows are made of plain glass, so the attenuation offered to the impinging wireless signal is negligible \cite{honcharenko1993mechanisms}. 


For an arbitrary Rx location, the path followed by a particular MPC can be represented as a string such as `Tx-X-X-$\cdots$-X-Rx'. Since the MPC begins at the transmitter and terminates at the receiver, all possible strings start with a `Tx' and terminate with an `Rx'. The character `X' in the string can be either `R' denoting `Reflection', `D' denoting `Diffraction' or `T' denoting `Transmission'. By reflection, we mean reflection from flat smooth surfaces following Snell's laws, transmission does not change the propagation direction and only attenuates the signal whereas diffraction is from edges and follows the diffraction law as in eq. \eqref{eq_law_of_diffraction}. The number of characters `X' between the `Tx' and `Rx' represents the number of interactions, and the left to right order of the characters denotes the sequence of interactions.
\par
Since in our 3D positioning technique we use the first arrival path principle, we define the First Arrival Path between a fixed Tx and Rx as
\begin{definition}
\textbf{First Arrival Path}: The particular MPC with the smallest ToF $\tau_{min}$ and SNR above a threshold $T$ among all existing MPCs at that location.
\end{definition}

 For the SNR calculation, we assume an Rx processing gain of 20 dB which is added to the signal power for each MPC extracted from the ray tracing simulation. Assuming a signal bandwidth of $B=400$ MHz and system temperature of $T=300$K, we obtain the noise floor as $N_0= KTB= -87$dBm. The threshold changes according to every pair of Rx-Tx according to $T=P_{max}-15$  where $P_{max}$ is the maximum SNR across all existing MPCs at a particular Rx location for a given Tx. 
 Now we conduct statistical analysis on the MPCs generated by the raytracing simulation across all UAV transmitters and receivers. First, we look at the elevation AoA and SNR for the first arrival path.

\begin{figure}[hpt]
    \centering
    \subfigure[Elevation AoA Boxplot.]{ 
        \begin{minipage}[t]{0.48\linewidth}
        \centering
        \includegraphics[clip, trim = 7cm 10cm 10cm 10cm, width=0.7\linewidth]{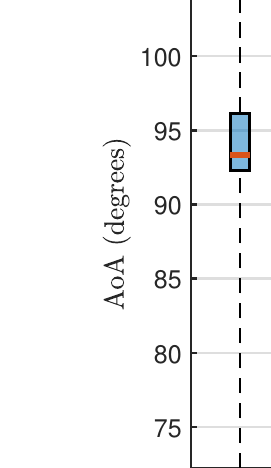}
        \end{minipage}}
    \subfigure[SNR Boxplot.]{
        \begin{minipage}[t]{0.48\linewidth}
        \centering
        \includegraphics[clip, trim = 7cm 10cm 10cm 10cm, width=0.7\linewidth]{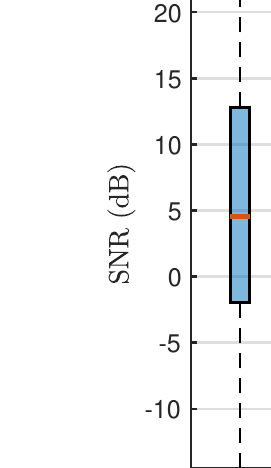}
        \end{minipage}}

        \caption{First arriving path statistics across Rx locations. We can see that the elevation AoA is always within six degrees of $90$\textdegree{} and that the SNR is typically $>5$dB.} 
        \label{fig_fap_metrics}
 \end{figure}

\begin{table}[ht]
\caption{MPC Through The Windows.}
\label{table_window_MPC}
\centering
\begin{tabular}{|c|c|c|c|c|}
\hline
\textbf{Group}& \textbf{MPC-1}& \textbf{MPC-2} & \textbf{MPC-3} & \textbf{MPC-4}   \\
\hline
Prop. Mech.& Tx-Rx& Tx-R-Rx & \shortstack{Tx-D-Rx \\ Tx-D-T-Rx \\ Tx-D-T-T-Rx} & \shortstack{Tx-R-R-$\cdots$-Rx \\ Tx-D-R-$\cdots$-Rx} \\
\hline
$P_e$&$3.47\%$&$9.03\%$&$91.52\%$& $100\%$ \\
\hline
$P_{fap}$&$3.47\%$&$1.60\%$&$84.79\%$& $10.12\%$ \\
\hline
\end{tabular}
\end{table}

\par
For a fixed Rx location and fixed Tx (UAV anchor), we can group the MPCs in four unique groups: (a) MPC-1, (b) MPC-2, (c) MPC-3, and (d) MPC-4 based on the propagation mechanism and number of interactions associated with the group. Table \ref{table_window_MPC} shows the four possible MPC groups, the string in the second row representing the corresponding propagation mechanism. Now we evaluate two metrics across all receivers and UAV anchors (a) Probability of existence: $P_e$ of the different MPC groups and (b) Probability that the first arrival path belongs to each MPC group: $P_{fap}$. Mathematically we can express $P_e$ and $P_{fap}$ as 
\begin{equation}
\begin{split}
    P_e = \sum_{tx}\frac{N_{e,tx}\times 100}{N_{rx}} ,\;\;\; P_{fap} = \sum_{tx} \frac{N_{fap,tx}\times 100}{N_{rx}}.
\end{split}
\end{equation}
Here, $N_{e,tx}$ is the number of Rx locations for each transmitter (UAV anchor) where a particular MPC group exists, and $N_{fap,tx}$ is the number of Rx locations for each transmitter where the first arriving path belongs to a particular MPC Group. We present three key observations on the different MPC groups based on Table \ref{table_window_MPC} and Fig. \ref{fig_fap_metrics}. 

\begin{observation}
For MPC-1 and MPC-2, we have a low probability of existence $P_e$.
\end{observation}
From the associated string in Table \ref{table_window_MPC}, MPC-1 contains MPCs that follow the direct Euclidean path (LoS path) between the transmitter and receiver, whereas MPC-2 contains MPCs that undergo a single bounce satisfying Snell's laws before reaching the receiver. The low probability of existence can be attributed to the fact that both these MPC groups exist only for Rx locations close to the windows. 

\begin{observation}
\label{observation_MPC3}
For MPC-3 we have a high we have a high probability of existence $P_e$ and high probability that they are also first arriving paths since $P_{fap}=84.79\%$. 
\end{observation}

The results of the ray tracing simulation indicate that two diffraction paths are generated due to the upper and lower edges of the window on the receiver floor, consistent with the simplified building model described in Section \ref{sec_simplified_model_of_building}. The first arriving path belongs to MPC-3 with the highest probability ($P_{fap}=84.79\%$). This outcome aligns with expectations, as the diffraction field follows the shortest path, a principle used to derive the path length for diffraction MPCs based on Fermat's principle of least time in Lemma \ref{lemma_half_plane_path_length}. Furthermore, in Fig. \ref{fig_fap_metrics}, it can be observed that the first arriving path most likely corresponds to MPC-3, with an elevation AoA close to 90\textdegree{}. This observation supports Proposition \ref{prop_diffraction_mpc_aoa_90}, which states that diffraction MPCs arrive at the receiver parallel to the ground.

\begin{observation}
The first arriving path principle rarely contains MPCs belonging to MPC-4.
\end{observation}
Since these MPCs belonging to this group undergo several interactions with the window and reflections from the insides of the building, we see varying path lengths. 

\subsection{Jacobian of the path length $p_{j}(\bm{\alpha})$}
The entries in the Jacobian matrix $J$ in eq. \eqref{eq_Jacobian} can be derived by using the chain rule on the path length $p_j(\bm{\alpha})$ between the $j^{th}$ anchor to the node at $\bm{\alpha}$ given by eq. \eqref{eq_path_length_building}. For each anchor, we obtain the x-coordinate $q_x$ of the diffraction point $Q_e$ on the diffracting edge by the expressions in Lemma \ref{lemma_half_plane_path_length} with the coefficients of the quadratic equation $a,b,c$ defined in eq. \eqref{eq_quadratic_qe} for each anchor with $z_e=z_n+0.5w$. For notational simplicity, drop the subscript $j$, and argument $\bm{\alpha}$ in the expression for the entries in the Jacobian. Hence, the path length is denoted as $p$ and is defined for the anchor at $\bm{X_a} = [x_a,y_a,z_a]^T$, the node at $\bm{\alpha}=[x_n,y_n,z_n]^T$ and the window size is $w$. The final expression is below.  

\footnotesize{
\begin{equation}
\label{eq_jacobian_elements}
\begin{split}
\frac{\partial p}{\partial x_n} & = \frac{(q_{x}-x_{a})\frac{\partial q_{x}}{\partial x_n}}{\sqrt{(q_{x}-x_{a})^2+(y_{a})^2+(z_{a}-z_{n}-0.5w)^2}} \\&\;\;\;\;\;\;\;\;\;\;\;\;\;\;\;\;\;\;\;\;\;\;\;+ \frac{(x_n-q_{x})(1-\frac{\partial q_{x}}{\partial x_n})}{\sqrt{(x_n-q_{x})^2+(y_{n})^2+(0.5w)^2}}, \\ 
\frac{\partial p}{\partial y_n} &= \frac{(q_{x}-x_{a})\frac{\partial q_{x}}{\partial y_n}}{\sqrt{(q_{x}-x_{a})^2+(y_{a})^2+(z_{a}-z_{n}-0.5w)^2}}\\& \;\;\;\;\;\;\;\;\;\;\;\;\;\;\;\;\;\;\;\; + \frac{(q_x-x_n)\frac{\partial q_x}{\partial y_n}+y_n}{\sqrt{(x_n-q_x)^2+(y_{n})^2+(0.5w)^2}}, \\
\frac{\partial p}{\partial z_n} &= \frac{(q_{x}-x_{a})\frac{\partial q_{x}}{\partial z_n}}{\sqrt{(q_{x}-x_{a})^2+(y_{a})^2+(z_{a}-z_{n}-0.5w)^2}}\\& \;\;\;\;\;\;\;\;\;\;\;\;\;\;\;\;\;\;\;\; + \frac{(q_x-x_n)\frac{\partial q_x}{\partial z_n}}{\sqrt{(x_n-q_x)^2+(y_{n})^2+(0.5w)^2}},  \\
\frac{\partial q_x}{\partial x_n} &= \frac{(x_1-x_2)}{2a}\left[\frac{\partial a}{\partial x_n}\left(\frac{(b \mp \sqrt{b^2-4ac}}{a}\right)\right.\\&\left.\;\;\;\;\;\;\;\;\;\;\;\;+\left(-\frac{\partial b}{\partial x_n}\pm \frac{b\frac{\partial b}{\partial x_n}-2c\frac{\partial a}{\partial x_n}-2a\frac{\partial c}{\partial x_n}}{\sqrt{b^2-4ac}}\right)\right], \\
\frac{\partial q_x}{\partial y_n} &= \frac{(x_1-x_2)}{2a}\left[\frac{\partial a}{\partial y_n}\left(\frac{(b \mp \sqrt{b^2-4ac}}{a}\right)\right.\\&\left.\;\;\;\;\;\;\;\;\;\;\;\;+\left(-\frac{\partial b}{\partial y_n}\pm \frac{b\frac{\partial b}{\partial y_n}-2c\frac{\partial a}{\partial y_n}-2a\frac{\partial c}{\partial y_n}}{\sqrt{b^2-4ac}}\right)\right], \\
\frac{\partial q_x}{\partial z_n} &= \frac{(x_1-x_2)}{2a}\left[\frac{\partial a}{\partial z_n}\left(\frac{(b \mp \sqrt{b^2-4ac}}{a}\right)\right.\\&\left.\;\;\;\;\;\;\;\;\;\;\;\;+\left(-\frac{\partial b}{\partial z_n}\pm \frac{b\frac{\partial b}{\partial z_n}-2c\frac{\partial a}{\partial z_n}-2a\frac{\partial c}{\partial z_n}}{\sqrt{b^2-4ac}}\right)\right], \\
\frac{\partial a}{\partial x_n} & = 0, \;
\frac{\partial a}{\partial y_n}  = 2y_n(x_1-x_2)^2,
\frac{\partial a}{\partial z_n}  = 2(x_1-x_2)^2(z_n+\frac{w}{2}),   \\
\frac{\partial b}{\partial x_n} & = 2(x_1-x_2)\left((z_1-z_{a})^2+y_{a}^2\right),\nonumber \\
\end{split}
\end{equation}
\begin{equation}
\begin{split}
\frac{\partial b}{\partial y_n}  &= 4y_n(x_1-x_2)(x_2-x_{a}), \\
\frac{\partial b}{\partial z_n} &= -4(x_1-x_2)(x_2-x_n)(z_n+\frac{w}{2}-z_a),\\
\frac{\partial c}{\partial x_n} & = 2(x_2-x_n)\left((z_1-z_{a})^2+y_{a}^2\right),\\
\frac{\partial c}{\partial y_n}  &= 2y_n(x_2-x_{a})^2 ,
\frac{\partial c}{\partial z_n}  = -2(x_2-x_n)^2(z_n+\frac{w}{2}-z_a).
\end{split}
\end{equation}
}

\bibliographystyle{IEEEtran}{
\footnotesize
\bibliography{refs}

\begin{thebibliography}{10}
\providecommand{\url}[1]{#1}
\csname url@samestyle\endcsname
\providecommand{\newblock}{\relax}
\providecommand{\bibinfo}[2]{#2}
\providecommand{\BIBentrySTDinterwordspacing}{\spaceskip=0pt\relax}
\providecommand{\BIBentryALTinterwordstretchfactor}{4}
\providecommand{\BIBentryALTinterwordspacing}{\spaceskip=\fontdimen2\font plus
\BIBentryALTinterwordstretchfactor\fontdimen3\font minus \fontdimen4\font\relax}
\providecommand{\BIBforeignlanguage}[2]{{%
\expandafter\ifx\csname l@#1\endcsname\relax
\typeout{** WARNING: IEEEtran.bst: No hyphenation pattern has been}%
\typeout{** loaded for the language `#1'. Using the pattern for}%
\typeout{** the default language instead.}%
\else
\language=\csname l@#1\endcsname
\fi
#2}}
\providecommand{\BIBdecl}{\relax}
\BIBdecl

\bibitem{duggal20243d}
G.~Duggal, R.~M. Buehrer, H.~S. Dhillon, and J.~H. Reed, ``{3D Positioning using a new diffraction path model},'' \emph{Proc., IEEE Intl. Conf. on Commun. (ICC)}, 2024.

\bibitem{kaplan2017understanding}
E.~D. Kaplan and C.~Hegarty, \emph{Understanding GPS/GNSS: principles and applications}.\hskip 1em plus 0.5em minus 0.4em\relax Artech house, 2017.

\bibitem{dwivedi2021positioning}
S.~Dwivedi, R.~Shreevastav, F.~Munier, J.~Nygren, I.~Siomina, Y.~Lyazidi, D.~Shrestha, G.~Lindmark, P.~Ernstr{\"o}m, E.~Stare \emph{et~al.}, ``Positioning in {5G} networks,'' \emph{IEEE Commun. Magazine}, Nov 2021.

\bibitem{shen2010fundamental}
Y.~Shen and M.~Z. Win, ``{Fundamental limits of wideband localization—Part I: A general framework},'' \emph{IEEE Trans. on Info. Theory}, 2010.

\bibitem{lonenlosbias}
C.~E. O’Lone, H.~S. Dhillon, and R.~M. Buehrer, ``{Characterizing the first-arriving multipath component in 5G millimeter wave networks: TOA, AOA, and non-line-of-sight bias},'' \emph{IEEE Trans. on Wireless Commun.}, 2022.

\bibitem{qi2006analysis}
Y.~Qi, H.~Kobayashi, and H.~Suda, ``{Analysis of wireless geolocation in a non-line-of-sight environment},'' \emph{IEEE Trans. on Wireless Commun.}, 2006.

\bibitem{jia2010set}
T.~Jia and R.~M. Buehrer, ``A set-theoretic approach to collaborative position location for wireless networks,'' \emph{IEEE Trans. Mobile Computing}, Dec 2010.

\bibitem{jiacol}
------, ``Collaborative position location with {NLOS} mitigation,'' \emph{Proc., IEEE PIMRC}, Dec 2010.

\bibitem{venkatesh2007nlos}
S.~Venkatesh and R.~M. Buehrer, ``{NLOS mitigation using linear programming in ultrawideband location-aware networks},'' \emph{IEEE Trans. on Veh. Technology}, 2007.

\bibitem{venkatesh2007non}
S.~Venkatesh and R.~Buehrer, ``Non-line-of-sight identification in ultra-wideband systems based on received signal statistics,'' \emph{IET Microwaves, Antennas \& Propagation}, Dec 2007.

\bibitem{li2023iterative}
Z.~Li, F.~Jiang, H.~Wymeersch, and F.~Wen, ``An iterative 5g positioning and synchronization algorithm in nlos environments with multi-bounce paths,'' \emph{IEEE Wireless Commun. Letters}, 2023.

\bibitem{Witrisalet_multipath_indoor_assisted}
K.~Witrisal, P.~Meissner, E.~Leitinger, Y.~Shen, C.~Gustafson, F.~Tufvesson, K.~Haneda, D.~Dardari, A.~F. Molisch, A.~Conti, and M.~Z. Win, ``High-accuracy localization for assisted living: {5G} systems will turn multipath channels from foe to friend,'' \emph{IEEE Signal Processing Magazine}, Mar 2016.

\bibitem{leitinger2015evaluation}
E.~Leitinger, P.~Meissner, C.~R{\"u}disser, G.~Dumphart, and K.~Witrisal, ``Evaluation of position-related information in multipath components for indoor positioning,'' \emph{IEEE Journal on Sel. Areas in Commun.}, Nov 2015.

\bibitem{mendrzik_nlos_mpc}
R.~Mendrzik, H.~Wymeersch, G.~Bauch, and Z.~Abu-Shaban, ``{Harnessing NLOS components for position and orientation estimation in 5G millimeter wave MIMO},'' \emph{IEEE Trans. on Wireless Commun.}, Dec 2019.

\bibitem{Hassanslam}
H.~Naseri and V.~Koivunen, ``Cooperative simultaneous localization and mapping by exploiting multipath propagation,'' \emph{IEEE Trans. on Signal Processing}, 2017.

\bibitem{Christianslam}
C.~Gentner, T.~Jost, W.~Wang, S.~Zhang, A.~Dammann, and U.-C. Fiebig, ``Multipath assisted positioning with simultaneous localization and mapping,'' \emph{IEEE Trans. on Wireless Commun.}, 2016.

\bibitem{Nazari6DSLAM}
M.~A. Nazari, G.~Seco-Granados, P.~Johannisson, and H.~Wymeersch, ``{MmWave 6D radio localization With a snapshot observation from a single BS},'' \emph{IEEE Trans. on Veh. Technology}, 2023.

\bibitem{leitinger_slam}
E.~Leitinger, A.~Venus, B.~Teague, and F.~Meyer, ``Data fusion for multipath-based slam: Combining information from multiple propagation paths,'' \emph{IEEE Trans. on Signal Processing}, 2023.

\bibitem{leitinger2019belief}
E.~Leitinger, F.~Meyer, F.~Hlawatsch, K.~Witrisal, F.~Tufvesson, and M.~Z. Win, ``A belief propagation algorithm for multipath-based slam,'' \emph{IEEE Trans. on Wireless Commun.}, 2019.

\bibitem{venus2023graph}
A.~Venus, E.~Leitinger, S.~Tertinek, and K.~Witrisal, ``A graph-based algorithm for robust sequential localization exploiting multipath for obstructed-los-bias mitigation,'' \emph{IEEE Trans. on Wireless Commun.}, 2023.

\bibitem{Firstnetroadmap}
\BIBentryALTinterwordspacing
{FirstNet Authority}. (2023, Sep) First responder network authority roadmap. [Online]. Available: \url{https://www.firstnet.gov/sites/default/files/Roadmap_2023.pdf}
\BIBentrySTDinterwordspacing

\bibitem{harishetal}
H.~K. Dureppagari, D.-R. Emenonye, H.~S. Dhillon, and R.~M. Buehrer, ``{UAV-aided indoor localization of emergency response personnel},'' in \emph{2023 IEEE/ION PLANS}, 2023.

\bibitem{duggaletal}
G.~Duggal, R.~M. Buehrer, N.~Tripathi, and J.~H. Reed, ``{Line-of-sight probability for outdoor-to-indoor UAV-assisted emergency networks},'' \emph{Proc., IEEE Intl. Conf. on Commun. (ICC)}, May 2023.

\bibitem{albaneseetal}
A.~Albanese, V.~Sciancalepore, and X.~Costa-Pérez, ``{First responders got wings: UAVs to the rescue of localization pperations in beyond 5G systems},'' \emph{IEEE Commun. Magazine}, 2021.

\bibitem{balanis2012advanced}
C.~A. Balanis, \emph{Advanced Engineering Electromagnetics}.\hskip 1em plus 0.5em minus 0.4em\relax John Wiley \& Sons, 2012.

\bibitem{namara1990introduction}
D.~Namara, C.~Pistorious, and J.~Maherbe, \emph{Introduction to the Uniform Geometric Theory of Diffraction}.\hskip 1em plus 0.5em minus 0.4em\relax Artech House, 1990.

\bibitem{bostian28}
P.~Tenerelli and C.~Bostian, ``{Measurements of 28 GHz diffraction loss by building corners},'' in \emph{Proc., IEEE PIMRC}, 1998.

\bibitem{honcharenko1993mechanisms}
W.~Honcharenko, H.~Bertoni, and J.~Dailing, ``Mechanisms governing propagation between different floors in buildings,'' \emph{IEEE Trans. on Antennas and Propagation}, 1993.

\bibitem{anurag_radarconfgtd}
A.~Pallaprolu, B.~Korany, and Y.~Mostofi, ``Analysis of {Keller} cones for rf imaging,'' in \emph{2023 IEEE Radar Conference (RadarConf23)}, 2023.

\bibitem{pallaprolu2022wiffract}
------, ``Wiffract: a new foundation for rf imaging via edge tracing,'' in \emph{Proc. of the 28th Annual Intl. Conf. on Mob. Computing And Networking}, 2022.

\bibitem{wirelessinsight}
\BIBentryALTinterwordspacing
{Remcom Inc }. (2024, July) {Wireless InSite MIMO: version 3.4.4}. [Online]. Available: \url{https://www.remcom.com}
\BIBentrySTDinterwordspacing

\bibitem{keller1962geometrical}
J.~B. Keller, ``Geometrical theory of diffraction,'' \emph{J. Opt. Soc. Am.}, 1962.

\bibitem{pathak_gtd}
P.~H. Pathak, G.~Carluccio, and M.~Albani, ``The uniform geometrical theory of diffraction and some of its applications,'' \emph{IEEE Antennas and Propagation Magazine}, 2013.

\bibitem{born2013principles}
M.~Born and E.~Wolf, \emph{Principles of optics: electromagnetic theory of propagation, interference and diffraction of light}.\hskip 1em plus 0.5em minus 0.4em\relax Elsevier, 2013.

\bibitem{rahmat2007keller}
Y.~Rahmat-Samfi, ``Keller's cone encountered at a hotel,'' \emph{IEEE Antennas and Propag. Mag.}, 2007.

\bibitem{kouyoumjian1974uniform}
R.~G. Kouyoumjian and P.~H. Pathak, ``A uniform geometrical theory of diffraction for an edge in a perfectly conducting surface,'' \emph{Proceedings of the IEEE}, 1974.

\bibitem{kohlio2i}
M.~Kohli, A.~Adhikari, G.~Avci, S.~Brent, A.~Dash, J.~Moser, S.~Hossain, I.~Kadota, C.~Garland, S.~Mukherjee, R.~Feick, D.~Chizhik, J.~Du, R.~A. Valenzuela, and G.~Zussman, ``{Outdoor-to-Indoor 28 GHz wireless weasurements in Manhattan: path loss, environmental effects, and 90\% coverage},'' \emph{IEEE/ACM Trans. on Networking}, 2024.

\bibitem{bas2019outdoor}
C.~U. Bas, R.~Wang, S.~Sangodoyin, T.~Choi, S.~Hur, K.~Whang, J.~Park, C.~J. Zhang, and A.~F. Molisch, ``{Outdoor to indoor propagation channel measurements at 28 GHz},'' \emph{IEEE Trans. on Wireless Commun.}, 2019.

\bibitem{siktaetal}
F.~Sikta, W.~Burnside, T.-T. Chu, and L.~Peters, ``First-order equivalent current and corner diffraction scattering from flat plate structures,'' \emph{IEEE Trans. on Antennas and Propagation}, 1983.

\bibitem{heath2016overview}
R.~W. Heath, N.~Gonzalez-Prelcic, S.~Rangan, W.~Roh, and A.~M. Sayeed, ``{An overview of signal processing techniques for millimeter wave MIMO systems},'' \emph{IEEE Journal on Sel. Areas in Signal Processing}, Feb 2016.

\bibitem{falsi2006time}
C.~Falsi, D.~Dardari, L.~Mucchi, and M.~Z. Win, ``{Time of arrival estimation for UWB localizers in realistic environments},'' \emph{EURASIP Journal on Adv. in Sig. Process.}, 2006.

\bibitem{van2004detection}
H.~L. Van~Trees, \emph{{Detection, Estimation, and Modulation Theory, part I: Detection, Estimation, and Linear Modulation Theory}}.\hskip 1em plus 0.5em minus 0.4em\relax John Wiley \& Sons, 2004.

\bibitem{kay1993fundamentals}
S.~M. Kay, \emph{Fundamentals of Statistical Signal Processing: Estimation Theory}.\hskip 1em plus 0.5em minus 0.4em\relax Prentice-Hall, Inc., 1993.

\bibitem{zekavat2011handbook}
R.~Zekavat and R.~M. Buehrer, \emph{{Handbook of Position Location: Theory, Practice and Advances}}.\hskip 1em plus 0.5em minus 0.4em\relax John Wiley \& Sons, 2011.

\bibitem{don_ris_near_far_field}
D.-R. Emenonye, H.~S. Dhillon, and R.~M. Buehrer, ``{RIS-aided localization under position and orientation offsets in the near and far field},'' \emph{IEEE Trans. on Wireless Commun.}, May 2023.

\bibitem{strang2020linear}
G.~Strang, \emph{{Linear Algebra for Everyone}}.\hskip 1em plus 0.5em minus 0.4em\relax SIAM, 2020.

\end{thebibliography}
}

\begin{IEEEbiography}[{\includegraphics[width=1in,height=1.25in,clip,keepaspectratio]{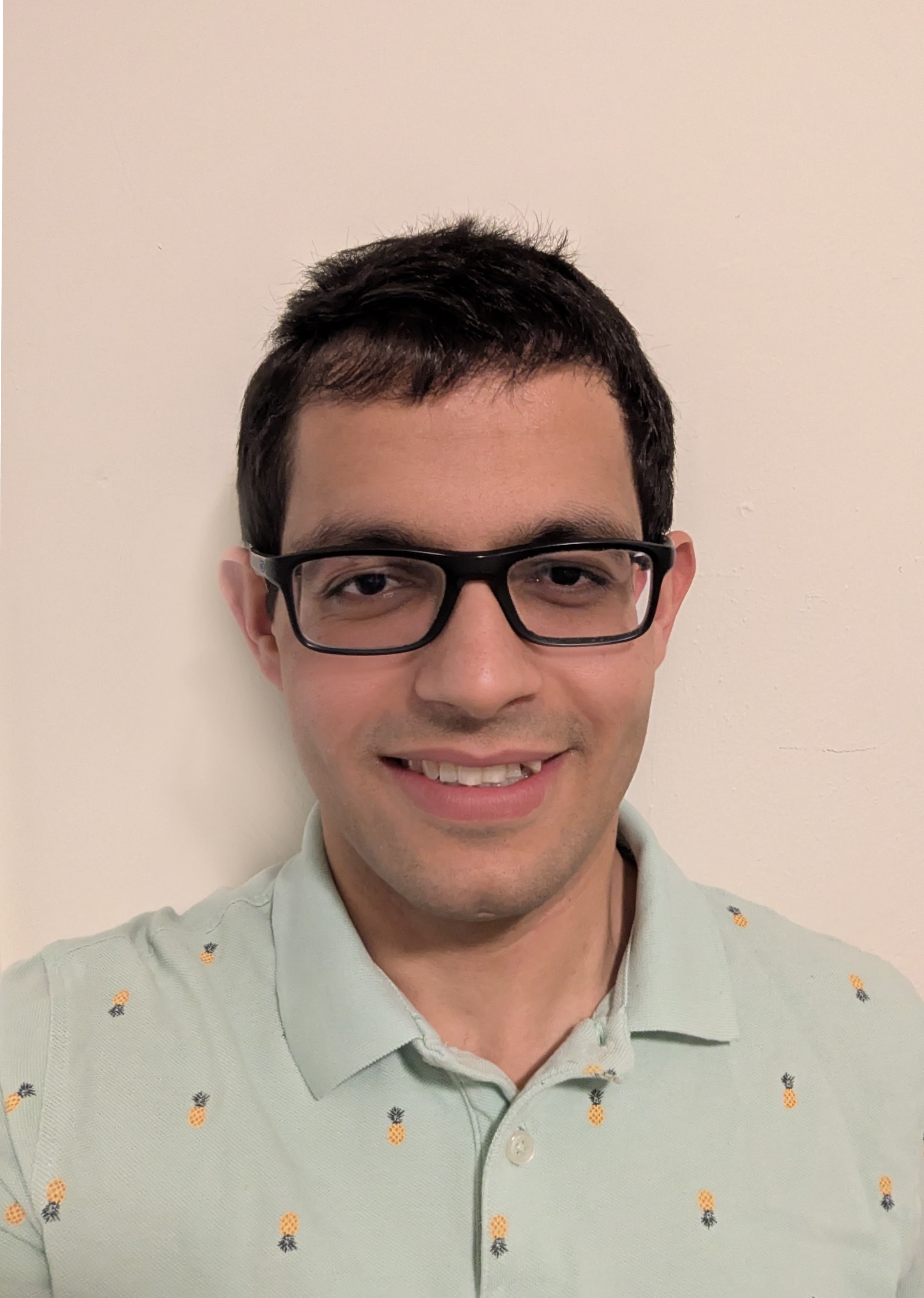}}]
{Gaurav Duggal} (Graduate Student Member, IEEE) received his B.E. degree in Electrical and Electronics Engineering from BITS Hyderabad in 2013 and his M.Tech. degree in Communications and Signal Processing from IIIT Delhi in 2019. He is currently a Ph.D. candidate with the Wireless@VT research group in the Bradley Department of Electrical and Computer Engineering at Virginia Tech. From 2019 to 2021, he worked as an Engineer at Qualcomm Technologies Inc., Hyderabad, India. He is a recipient of the NIJ FY23 Graduate Research Fellowship (2023–2026). His research interests include wireless communications, wireless localization, and radar systems.
\end{IEEEbiography}

\begin{IEEEbiography} [{\includegraphics[width=1in,height=1.25in,clip,keepaspectratio]{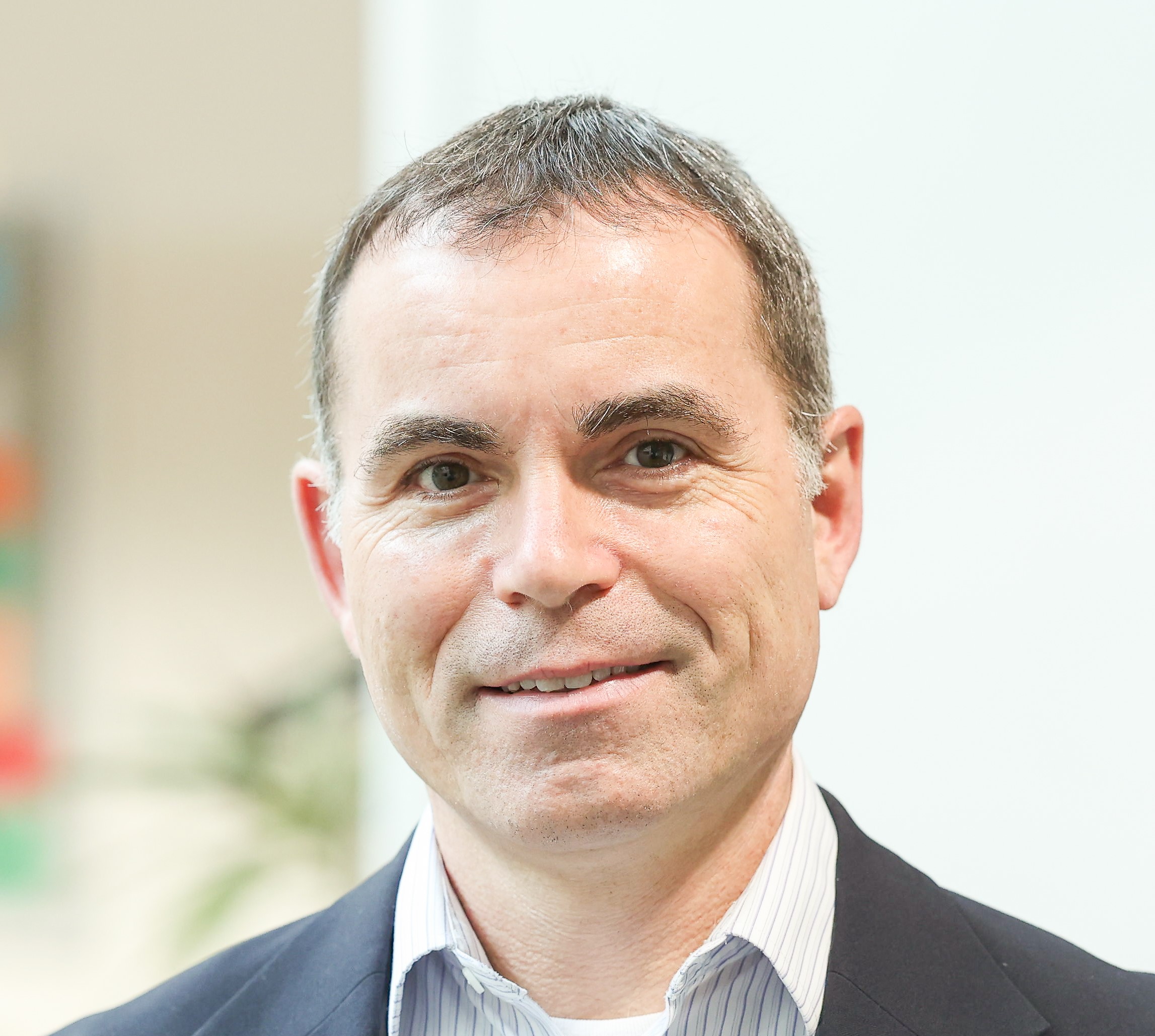}}]{R. Michael Buehrer}(Fellow, IEEE) joined Virginia Tech from Bell Labs as an Assistant Professor with the Bradley Department of Electrical and Computer Engineering in 2001.  He is currently a Professor of Electrical Engineering and is the Director of Wireless @ Virginia Tech, a comprehensive research group focusing on wireless communications, radar and localization.  During 2009 Dr. Buehrer was a visiting researcher at the Laboratory for Telecommunication Sciences (LTS) a federal research lab which focuses on telecommunication challenges for national defense.  While at LTS, his research focus was in the area of cognitive radio with a particular emphasis on statistical learning techniques.
Dr. Buehrer was named an IEEE Fellow in 2016 “for contributions to wideband signal processing in communications and geolocation.”  In 2023, he received the prestigious MILCOM Lifetime Award for Technical Achievement. This award recognizes individuals who have made important technical contributions to military communications over the course of their careers. His current research interests include machine learning for wireless communications and radar, geolocation, position location networks, cognitive radio, cognitive radar, electronic warfare, dynamic spectrum sharing, communication theory, Multiple Input Multiple Output (MIMO) communications, spread spectrum, interference avoidance, and propagation modeling.  His work has been funded by the National Science Foundation, the Defense Advanced Research Projects Agency, the Office of Naval Research, the Army Research Office, the Air Force Research Lab and several industrial sponsors.
Dr. Buehrer has authored or co-authored over 100 journal and approximately 275 conference papers and holds 18 patents in the area of wireless communications.  In 2023 and 2021 he was the co-recipient of the Vanu Bose Award for the best paper at MILCOM’21.  In 2023 and 2010 he was co-recipient of the Fred W. Ellersick MILCOM Award for the best paper in the unclassified technical program.  He was formerly an Area Editor IEEE Wireless Communications.  He was also formerly an associate editor for IEEE Transactions on Communications, IEEE Transactions on Vehicular Technologies, IEEE Transactions on Wireless Communications, IEEE Transactions on Signal Processing, IEEE Wireless Communications Letters, and IEEE Transactions on Education.  He has also served as a guest editor for special issues of The Proceedings of the IEEE, and IEEE Transactions on Special Topics in Signal Processing.  In 2003 he was named Outstanding New Assistant Professor by the Virginia Tech College of Engineering and in 2014 he received the Dean’s Award for Excellence in Teaching.
\end{IEEEbiography}

\begin{IEEEbiography} [{\includegraphics[width=1in,height=1.25in,clip,keepaspectratio]{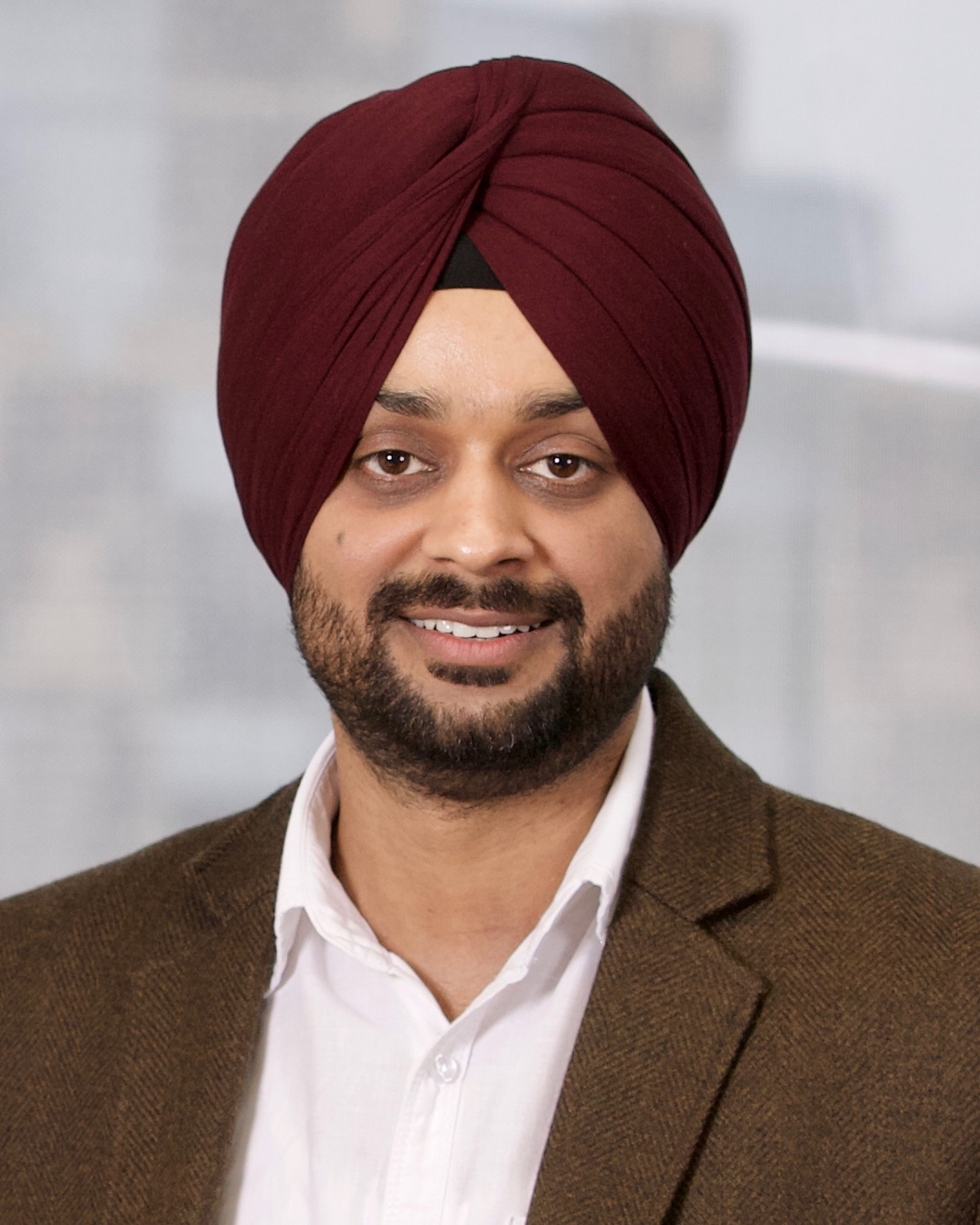}}]{Harpreet S. Dhillon}               
(Fellow, IEEE) received his B.Tech. from IIT Guwahati in 2008, his M.S. from Virginia Tech in 2010, and his Ph.D. from the University of Texas at Austin in 2013. After a year as a Viterbi Postdoctoral Fellow at the University of Southern California, he joined Virginia Tech in 2014, where he is currently the W. Martin Johnson Professor of Engineering and the Associate Dean for Research and Innovation. In 2024, he served as the Interim Department Head of Electrical and Computer Engineering (ECE). His research interests include communication theory, wireless networks, geolocation, and stochastic geometry. He is a Fellow of AAIA and AIIA, and a Clarivate Analytics Highly Cited Researcher. He has received six best paper awards, including the IEEE Leonard G. Abraham Prize (2014), the IEEE ComSoc Young Author Best Paper Award (2015), and the IEEE Heinrich Hertz Award (2016), as well as early-career technical achievement awards from three IEEE ComSoc Technical Committees. At Virginia Tech, he has received multiple faculty fellowships and research excellence awards, including the Outstanding New Assistant Professor Award (2017), the Dean’s Award for Excellence in Research (2020), and named fellowships such as the Steven O. Lane Junior Faculty Fellow (2018), the College of Engineering Faculty Fellow (2018), and the Elizabeth and James E. Turner Jr. ’56 Faculty Fellow (2019). His other honors include the 2008 Agilent Engineering and Technology Award, the UT Austin MCD Fellowship, the 2013 UT Austin WNCG Leadership Award, and the inaugural IIT Guwahati Young Alumni Achiever Award (2020). He has served as a TPC Co-chair for IEEE WCNC 2022 and IEEE PIMRC 2024, a symposium TPC Co-chair for numerous IEEE conferences, and as an editorial board member of several IEEE journals. He currently serves on the Executive Editorial Committee for the {\sc IEEE Transactions on Wireless Communications}.
\end{IEEEbiography}

\begin{IEEEbiography} [{\includegraphics[width=1in,height=1.25in,clip,keepaspectratio]{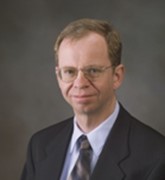}}]{Jeffrey Reed}(Fellow, IEEE) is the Willis G. Worcester Professor of ECE. Professor Reed’s research interests are wireless communications, wireless security, cognitive radio, software radio, telecommunications policy, and spectrum access. Reed has co-authored more than 500 articles and books. In addition, Reed co-founded several commercial companies, including Federated Wireless, which commercializes spectrum sharing; PFP Cybersecurity, which provides security solutions for IoT devices; and Cirrus360, which produces tools for rapid prototyping of O-RAN. Reed is the Founding Director of Wireless@Virginia Tech, a university research center, and co-founder of Virginia Tech’s Hume Center for National Security and Technology, where he served as the Interim Director. He also served as the Interim Director of the Commonwealth Cyber Initiative and is currently its CTO. Dr. Reed is a Fellow of the IEEE For contributions to software radio and communications signal processing and for leadership in engineering education.
\end{IEEEbiography}

\end{document}